\documentclass[twocolumn,twocolappendix]{aastex701}

\usepackage{xcolor}
\usepackage[most]{tcolorbox}
\usepackage{subcaption}
\usepackage[bottom]{footmisc}
\usepackage{booktabs}

\tcbuselibrary{skins,breakable}
\newtcolorbox{mybox}[2][]{breakable,sharp corners, skin=enhancedmiddle jigsaw,parbox=false,
boxrule=0mm,leftrule=2mm,boxsep=0mm,arc=0mm,outer arc=0mm,attach title to upper,
after title={\\ }, coltitle=black,colback=gray!10,colframe=black, title={#2},
fonttitle=\bfseries,#1}

\newcommand{\binc}{\texttt{binary\_c}}

\newcommand{\mesa}{\texttt{MESA}}

\newcommand{\npcebtot}{193}

\shorttitle{The Best Guess: Testing formalisms for the common envelope}
\shortauthors{Thai et al.}

\graphicspath{{./}}

\hypersetup{
	colorlinks=true,
	linkcolor=blue,
	citecolor=blue,
	urlcolor=cyan,
  pdftitle={The Best Guess (Thai et al., 2026)}
}
\sloppypar
\begin{document}

\title{The Best Guess: Testing new and old formalisms for the common envelope against observations}

\author[0009-0000-9368-0006]{Riley Thai}
\affiliation{School of Physics \& Astronomy, Monash University, Melbourne, VIC 3800, Australia}
\email[show]{riley.thai@monash.edu}

\author[0000-0002-3625-6951]{Amanda I. Karakas}
\affiliation{School of Physics \& Astronomy, Monash University, Melbourne, VIC 3800, Australia}
\email{amanda.karakas@monash.edu}

\author[0000-0001-5546-6869]{Zara Osborn}
\affiliation{School of Physics \& Astronomy, Monash University, Melbourne, VIC 3800, Australia}
\email{zara.osborn@monash.edu}

\author[0000-0003-0378-4843]{Robert G. Izzard}
\affiliation{School of Mathematics and Physics, University of Surrey, Guildford, GU2 7XH, UK}
\email{rob.izzard@surrey.edu}

\author[0000-0002-8032-8174]{Ryosuke Hirai}
\affiliation{School of Physics \& Astronomy, Monash University, Melbourne, VIC 3800, Australia}
\affiliation{RIKEN Cluster for Pioneering Research (CPR), RIKEN, Wako, Saitama 351-0198, Japan}
\email{ryosuke.hirai@monash.edu}

\author[0000-0003-2059-5841]{Alex J. Kemp}
\affiliation{Institute of Astronomy (IvS), KU Leuven, Celestijnenlaan 200D, B-3001 Leuven, Belgium}
\email{alex.kemp@monash.edu}

\author[0000-0001-5298-0694]{Simon W. Campbell}
\affiliation{School of Physics \& Astronomy, Monash University, Melbourne, VIC 3800, Australia}
\email{simon.campbell@monash.edu}

\begin{abstract}
	\noindent
	We present a systematic test of formalisms for common envelope evolution by forward-modelling observable post-common envelope binaries. We compare predictions from the $\alpha$-formalism, and the Two-stage and SCATTER formalisms against observed post-common envelope binaries, including wide binaries with ultra-massive white dwarfs and central binaries of planetary nebulae. The angular momentum-based SCATTER formalism does not predict populations which match the complete observed population, even with adjustments to its parameters. We take this as indicative of fundamental challenges with using the orbital angular momentum balance to predict common envelope outcomes. The energy-based $\alpha$ and hybrid Two-stage formalisms both well-replicate the observed population. $\alpha_{\mathrm{CE}} \sim 0.2\text{--}0.3$ can match current observations, in agreement with previous works. Recombination energy is necessary, but only a fraction of it ($\sim\! 10\text{--}40\%$) can contribute in order to predict IK Peg-like binaries with ultra-massive white dwarfs at the correct orbital periods. Our work suggests energy-based formalisms remain the most accurate for predicting common envelope outcomes, but more observations can constrain the recombination contribution and how these outcomes systematically vary with the donor mass.
\end{abstract}


\section{Introduction}

\label{sec:intro}
The evolutionary process which determines the formation of closely separated, evolved binaries is the \emph{common envelope} (CE) interaction \citep{webbinkEvolutionLowmassClose1975, paczynskiCommonEnvelopeBinaries1976, demarcoDawesReview62017,ropkeSimulationsCommonenvelopeEvolution2023}. It occurs when an evolved, giant star undergoes unstable Roche lobe overflow (RLOF), which unbinds the envelope to encircle both components. This \emph{common envelope} is then ejected, leaving a hardened binary or a single merged giant.
Common envelope events can lead to the close binaries responsible for Type Ia supernovae \citep[e.g.,][]{ruiterDelayTimesRates2011,claeysTheoreticalUncertaintiesType2014,yungelsonMergingWhiteDwarfs2017}, gravitational wave sources \citep[e.g.,][]{kruckowProgenitorsGravitationalWave2018}, stripped envelope supernovae \citep[e.g.,][]{clocchiattiLightCurvesStrippedEnvelope1997,souropanisPowerBinariesStrippedenvelope2026}, and closely separated, semi-detached binaries of various types \citep[e.g., AM CVn, sdOBA-type hot subdwarfs, central binaries in planetary nebulae, and cataclysmic variables;][]{roelofsPopulationAMCVn2007,heberHotSubdwarfStars2009,jonesBinaryStarsKey2017}.
However, the exact physics of common envelope evolution remains elusive \citep[e.g.,][]{ivanovaCommonEnvelopeEvolution2013,ropkeSimulationsCommonenvelopeEvolution2023}, and an accurate understanding of the physical processes is vital for precisely determining the evolutionary outcomes of these systems.

Predicting the outcome of a common envelope event is often done by considering the energy budget of the final and initial states of the binary \citep{webbinkDoubleWhiteDwarfs1984}. The orbital energy of the binary is used to overcome the envelope's binding energy, modulated by an efficiency parameter \( \alpha_{\mathrm{CE}}\) to account for additional sources or sinks, as

\begin{equation}
	E_{\mathrm{bind}} = \alpha_{\mathrm{CE}} \Delta E_{\mathrm{orb}} \label{eq:alpha}
	.\end{equation}

Equation \ref{eq:alpha} remains the most widely used prescription to predict common envelope outcomes, whether for detailed stellar evolution codes \citep{marchantRoleMassTransfer2021,fragosPOSYDONGeneralpurposePopulation2023} or rapid population synthesis \citep[e.g.,][]{yungelsonModelGalacticPopulation1995,nelemansPopulationSynthesisDouble2001a,hanOriginSubdwarfStars2002,izzardBinaryOriginLowluminosity2004,rileyRapidStellarBinary2022}.
However, neither observations nor simulations conclusively support this energy-based `\( \alpha  \)-formalism' as a complete description of the common envelope process.

Observations and population synthesis of short period post-common envelope binaries (PCEBs) (\( \lesssim\!3 \) days) can constrain \( \alpha_{\mathrm{CE}}  \) to \( 0.2\text{--}0.3 \) \citep[e.g.,][]{zorotovicPostcommonenvelopeBinariesSDSS2010,toonenEffectCommonenvelopeEvolution2013, hernandezWhiteDwarfBinary2021,hernandezWhiteDwarfBinary2022,hernandezWhiteDwarfBinary2022a,geCommonEnvelopeEvolution2024,santos-garciaPopulationSynthesisStudy2025}. Longer period (\( \gtrsim\!10 \) days) binaries conflictingly suggest near unity efficiencies or otherwise a contribution from an additional energy source such as recombination energy \citep[e.g.,][]{nelemansPopulationSynthesisDouble2001,nelemansReconstructingEvolutionWhite2005,davisComprehensivePopulationSynthesis2010,zorotovicEvolutionSelflensingBinary2014,yamaguchiWidePostcommonEnvelope2024, yamaguchiWidePostcommonEnvelope2024a, belloniFormationLongperiodPostcommon2024}.
Low values for \( \alpha_{\mathrm{CE}}  \) also typically overpredict observed event rates \citep[e.g.,][]{ruiterDelayTimesRates2011,yungelsonMergingWhiteDwarfs2017,kempImpactMetallicityNova2022} and the number of semi-detached systems \citep[e.g., AM CVn,][]{nelemansShortperiodAMCVn2004,heberHotSubdwarfStars2009}.
The population synthesis results also sensitively depend on the parametrization used to determine the binding energy budget \citep{taurisResearchNoteBinding2001,ivanovaCommonEnvelopeEvolution2013,sgallettaImpactEnvelopeBinding2026} and boundaries of mass transfer stability \citep[e.g.,][]{liEvolutionWhiteDwarf1997,claeysTheoreticalUncertaintiesType2014,pavlovskiiStabilityMassTransfer2017,liInfluenceMassTransfer2023}.

Detailed three-dimensional hydrodynamical simulations of common envelope inspiral always fail to eject the envelope without the inclusion of recombination energy in their equation of state \citep[e.g.,][]{sandquistDoubleCoreEvolution1998,passySimulatingCommonEnvelope2012,nandezRecombinationEnergyDouble2015}. However, its overall effectiveness upon the final separation still remains debated and largely uncharacterized across different stellar evolutionary phases \citep[e.g.,][]{grichenerLimitedRoleRecombination2018,ivanovaUseHydrogenRecombination2018,iaconiSpeakingOneVoice2019,reichardtImpactRecombinationEnergy2020,gonzalez-bolivarCommonEnvelopeBinary2022,wilsonConvectionReconcilesDifference2022,lauCommonEnvelopesMassive2025}.

\begin{figure*}[t]
	\centering
	\includegraphics[width=\textwidth]{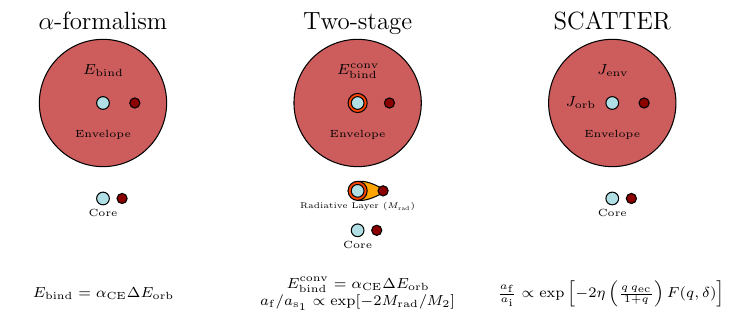}
	\caption{Schematic diagram of each formalism, along with the corresponding predictive method for the final state.
		\emph{Left}: the standard \( \alpha\)-formalism follows the energy budget available for the envelope's ejection. The orbital energy is used to overcome the binding energy of the envelope, modulated by an efficiency \( \alpha_{\mathrm{CE}} \) per Eq. \ref{eq:alpha} and \ref{eq:ebind}.
		\emph{Middle:} The hybrid Two-stage only assumes that the convective (isentropic) portion of envelope is ejected per energy conservation (stage I), and the ejection of the radiative layer above the outermost burning shell follows conservation of orbital angular momentum (stage II). We take this as a fully non-conservative process where mass is lost isotropically from the vicinity of the accretor, per Eq. \ref{eq:twostage}.
		\emph{Right}: the SCATTER formalism follows the orbital angular momentum budget, treating the case where the common envelope exerts a torque on each component (the core and companion). The final separation is proportional to an exponential term, as shown in Eq. \ref{eq:scatter}.}
	\label{fig:schematic}
\end{figure*}

With a lack of strong physical constraints, values of the efficiency \( \alpha_{\mathrm{CE}} \) are chosen arbitrarily in most binary population synthesis studies. Varying \( \alpha_{\mathrm{CE}}  \) and the parametrization of \( E_{\mathrm{bind}} \) to account for these uncertainties results in large uncertainties in the predictions from these models, especially for systems which undergo more than one common envelope event \citep[e.g.,][]{ruiterDelayTimesRates2011,claeysTheoreticalUncertaintiesType2014,yungelsonMergingWhiteDwarfs2017,kruckowProgenitorsGravitationalWave2018,osbornUsingBinaryPopulation2025}.

Recently, new formalisms of common envelope evolution have been published in the literature to attempt a better characterization of the underlying physics \citep[e.g.,][]{hiraiTwostageFormalismCommonenvelope2022,distefanoSCATTERNewCommon2023}.
The observed population of PCEBs has also been expanded significantly to higher companion masses (\( >\! 0.8 M_{\odot} \)) and orbital periods over 100 days \citep{hernandezWhiteDwarfBinary2021,hernandezWhiteDwarfBinary2022,hernandezWhiteDwarfBinary2022a,yamaguchiWidePostcommonEnvelope2024,yamaguchiWidePostcommonEnvelope2024a,shiraishiTwoUnseenMassive2026,motherwayNotsocompactCompanionMassive2026}. These formalisms are yet to be tested against the complete population of observed systems.

Here, we present the first systematic investigation of these new, alternative models for predicting common envelope outcomes, confronting their predictions with the largest compilation of observed white-dwarf main-sequence (WD-MS) PCEBs from the literature. We assess how the common envelope outcomes change under the Two-stage formalism \citep{hiraiTwostageFormalismCommonenvelope2022} and the SCATTER formalism \citep{distefanoSCATTERNewCommon2023} compared to the standard \( \alpha  \)-formalism.

This paper is organized as follows:
Section \ref{sec:methods} describes these new formalisms, our binary population synthesis model and its input physics.
Section \ref{sec:results} presents the key results of our models, while we discuss our results in Section \ref{sec:discussion}.
Finally, we summarize our findings in Section \ref{sec:conclusion}.
We additionally describe our development of parametrizations for the Two-stage formalism and provide the complete polynomial fits to the binding energy and radiative region mass in Appendix \ref{sec:twostage_apply}.

\section{Binary population model}
\label{sec:methods}
\subsection{Contemporary common envelope formalisms}
Here, we describe the two new common envelope formalisms that we test against observations. A schematic diagram of the processes for the two new formalisms is shown in Fig. \ref{fig:schematic}.
\subsubsection{The Two-stage formalism}
\label{sec:intro_twostage}
The Two-stage formalism \citep{hiraiTwostageFormalismCommonenvelope2022} is a modification of the \( \alpha  \)-formalism which accounts for the difference in the entropy of the stellar structure, considering both the energy and orbital angular momentum budget across distinct stages where each conservation law applies. The common envelope process is separated into two stages: an adiabatic plunge-in phase for the (nearly) isentropic convective envelope governed by energy balance, followed by a stable mass transfer phase for the inner radiative region where the entropy gradient is steep, which is governed by orbital angular momentum loss in stable but non-conservative mass transfer. The formalism is motivated by contemporary results from three-dimensional hydrodynamical simulations of the dynamical inspiral in massive stars \citep[e.g.,][]{fragosCompleteEvolutionNeutronstar2019,gonzalez-bolivarCommonEnvelopeBinary2022}, and the physical response of a striped donor's radiative region above the core \citep{marchantRoleMassTransfer2021,vigna-gomezStellarResponseStripping2022}. Fig. \ref{fig:schematic} shows a schematic diagram of the two stages in the central column.

The first stage is computed by the standard \( \alpha  \)-formalism given by Eq. \ref{eq:alpha}, except only considering the energy budget of the isentropic portion of the envelope. Assuming all material is transferred to the vicinity of the companion and ejected isotropically (i.e., isotropic re-emission), the final separation can be written as a function of the post-stage I separation \( a_{\mathrm{s_{1}}} \) \citep{postnovEvolutionCompactBinary2014}.
\begin{equation}
	\begin{split}
		\frac{a_{\mathrm{f}}}{a_{\mathrm{s_{1}}}} & = \frac{M_{\mathrm{tot}}}{M_{\mathrm{core}}+M_{2}}                                                                                          \\
		                                          & \times\left( \frac{M_{\mathrm{core}}+M_{\mathrm{rad}}}{M_{\mathrm{core}}} \right) ^2 \exp \left( -2 \frac{M_{\mathrm{rad}}}{M_{2}} \right),
	\end{split}\label{eq:twostage}
\end{equation}
where \( M_{\mathrm{rad}} \) is the mass of the radiative region, and \( M_{\mathrm{core}} \) is the mass of the core of the giant star.

While designed for massive stars, intermediate-mass (\( 2\text{--}8 M_{\odot} \)) giant stars have a similar entropy structure throughout their evolution up to the asymptotic giant branch (AGB), and the Two-stage formalism can apply for common envelope interactions prior to it. We find that the Two-stage formalism can apply to pre-AGB giant donors above \( M>2.25M_{\odot} \), approximately where stars have non-degenerate helium core ignition, and will thus predict differences in part of the WD-MS PCEB population.
Our full investigation of where the formalism applies and fits to \( M_{\mathrm{rad}} \) and the binding energy of convective envelope are provided in Appendix \ref{sec:twostage_apply}.

\subsubsection{The SCATTER formalism}
The Single Components' Angular momenTum TransfER (SCATTER) formalism \citep[][]{distefanoSCATTERNewCommon2023} utilizes the orbital angular momentum budget for its parametrization (right column, Fig. \ref{fig:schematic}). Whereas previous formalisms considering the orbital angular momentum budget assumed linear proportionalities \citep[i.e., the \( \gamma  \)-formalism;][]{nelemansReconstructingEvolutionDouble2000,nelemansReconstructingEvolutionWhite2005,webbinkCommonEnvelopeEvolution2008}, the SCATTER formalism is intended to take more general assumptions for use in higher order multiples \citep[see][]{khwajaGeneralizingSCATTERFormalism2025}. The formalism treats the case where the envelope drains angular momentum from each component interior to the envelope for its ejection. Under these assumptions, the final separation of a system can be predicted as
\begin{equation}
	\begin{split}
		\frac{a_{\mathrm{f}}}{a_{\mathrm{i}}} & = \left[ (1+q_{\mathrm{ec}})^2\frac{M_{\mathrm{core}}+M_{2}}{M_{\mathrm{tot}}} \right] \\
		                                      & \times \exp \left[
		-2 \eta  \left(  \frac{q_{\rm cc}\,r_{\rm L}(q_{\rm cc}^{-1} )^{\delta }  + q_{\rm cc}^{-1}r_{\rm L}(q_{\rm cc})^{\delta } }{r_{\rm L}(q_{\rm cc})^{\delta } + r_{\rm L}(q_{\rm cc}^{-1})^{\delta }  }\right) \right]
		,\end{split}\label{eq:scatter}
\end{equation}
for envelope-to-core mass ratio \( q_{\mathrm{ec}} = M_{\mathrm{env}} / M_{\mathrm{core}} \) and companion-to-core mass ratio \( q_{\mathrm{c c}}= M_{\mathrm{core}}/M_{2} \), also shown on the right of Fig. \ref{fig:schematic}. \( \eta  \) describes the proportionality between the change in the angular momentum of each companion and the specific angular momentum of the core for each amount of envelope mass with which it interacts. \( \delta  \) is a dimensionality parameter, which best constrains \( \eta  \) when \( \delta =3 \) \citep{distefanoSCATTERNewCommon2023}. The function \( r_{\mathrm{L}}(q)  \) is the Eggleton dimensionless Roche lobe parameter \citep{eggletonAproximationsRadiiRoche1983}. The exponent term describes how the mass ratio influences the amount of envelope mass that interacts with each component \citep[see Sec. 3.2,][]{distefanoSCATTERNewCommon2023}.

\subsection{Binary population synthesis}
\label{sec:popsynth}
To model our binary populations, we use \binc{} \citep{izzardBinaryOriginLowluminosity2004, izzardBinary_cStellarPopulation2023}. \binc{} is built on the foundations of \texttt{BSE} \citep{hurleyEvolutionBinaryStars2002}, but is purpose-built for modelling low-and-intermediate-mass stellar binaries and tracking nucleosynthetic yields.
We use a modified version of \binc{} version 2.2.4\footnote{The modified version of \binc{} used in this paper is available online at \href{https://gitlab.com/rileythai/binary_c/-/tree/twostage-v2.2.4?ref_type=heads}{gitlab.com/rileythai/binary\_c:twostage-v2.2.4}.}, interfaced with \texttt{binary\_c-python} version 1.0.0.

Our input physics prescriptions and parameters are described in Table \ref{tab:params}. We choose a series of state-of-the-art input physics designed to reflect our current understanding of the physics.
\binc{} includes
a more accurate interpolation and extrapolation of the \citet{polsStellarEvolutionModels1998} sequences using a logarithmic tabular interpolation \citep{schneiderEvolutionMassFunctions2015},
updated parametrizations of thermally-pulsing asymptotic giant branch (TPAGB) stellar structure based on the Monash models \citep{karakasParameterisingThirdDredgeup2002,izzardBinaryOriginLowluminosity2004,izzardNewSyntheticModel2004},
accurate treatment of accreting WD's for modelling individual novae events \citep{kempPopulationSynthesisAccreting2021},
and an updated treatment of RLOF \citep{claeysTheoreticalUncertaintiesType2014}.
For this work, we also implement the Two-stage formalism, update white dwarf cooling and evolutionary models \citep{althausNewEvolutionarySequences2013,camisassaUpdatedEvolutionarySequences2017,camisassaEvolutionUltramassiveWhite2019,camisassaWhiteDwarfStars2025},
update stability bounds for dynamical mass transfer \citep{geAdiabaticMassLoss2010,geAdiabaticMassLoss2015,geAdiabaticMassLoss2020,geThermalEquilibriumMassloss2020,zhangAdiabaticMassLoss2024},
and use an empirically-motivated disrupted and saturated magnetic braking prescription \citep{belloniEvidenceSaturatedDisrupted2024}.
We further compare varying prescriptions for binding energies, magnetic braking, and \( \alpha_{\mathrm{CE}} \) in Sec. \ref{sec:discussion}.

We assume fully circular orbits at formation \citep{hurleyEvolutionBinaryStars2002} with no initial stellar rotation, and sample for primary mass \( M_{1,0}\), initial mass ratio \( q_{0} \), and initial orbital period \( P_{0} \) (see Table \ref{tab:params}). We sample 80 equally spaced values, totalling to \( 80\times 80\times 80 = 512,000 \) binaries, and simulate all systems for 15 Gyr.
We adopt a solar metallicity composition (\( Z_{\odot}=0.014 \), \citealt{asplundChemicalCompositionSun2009}), as the majority of observed PCEBs have near Solar/thin disc metallicities. We discuss these assumptions further in Sec. \ref{sec:mh_ecc}.

\subsubsection{Common envelope formalisms}
For common envelope, we use the \( \alpha  \)-formalism, the Two-stage formalism, or the SCATTER formalism, per Eqs. \ref{eq:alpha}, \ref{eq:twostage} and \ref{eq:scatter} respectively.
Based on our investigation  (see Sec. \ref{sec:intro_twostage} and Appendix \ref{sec:twostage_apply}), we limit applying the Two-stage formalism for common envelope to donors in the Hertzsprung gap (HG), red giant branch (RGB), and core helium burning (CHeB) phases with \( M>\!2.25\,M_{\odot} \), and use the standard \( \alpha  \)-formalism for other stellar types and lower masses. For the SCATTER model, we apply it to all giant-like donors.

We adopt \( \alpha_{\mathrm{CE}} = 0.2 \) in our standard model given the constraints from previous studies \citep[e.g.,][]{zorotovicPostcommonenvelopeBinariesSDSS2010,demarcoFormalismCommonEnvelope2011}. We assume \( \alpha_{\mathrm{CE}} \) is identical for both the Two-stage formalism and \( \alpha  \)-formalism, as both involve an identical process of the ejection of the convective (isentropic) envelope. We use the functional form for proportionality parameter \( \eta  \) from \citet{distefanoSCATTERNewCommon2023},
\begin{equation}
	\log_{10} (\eta)  = 0.603 - 0.952 \cdot \log_{10}(q_{\mathrm{ec}})  \label{eq:etafit}
	,\end{equation}
in our standard model.

If a given system enters double-core common envelope (both stars are giant-like), we apply the standard \( \alpha  \)-formalism, as it is not clear how the assumptions imposed in the Two-stage's stage II mass transfer or the SCATTER formalism apply in this case.

\subsubsection{Envelope binding energies}
For the \( \alpha  \)-formalism, we parametrize the binding energy as \citep{dekoolCommonEnvelopeEvolution1990,hanFormationBipolarPlanetary1995},
\begin{equation}
	E_{\mathrm{bind}} = E_{\mathrm{grav}} + f_{\mathrm{th}}E_{\mathrm{th}} + f_{\mathrm{ion}}E_{\mathrm{rec}} = \frac{-GM_{\star}M_{\mathrm{env}}}{\lambda R}\label{eq:ebind}
	,\end{equation}
where \( M_{\star} \) is the total mass of the star, \( M_{\mathrm{env}} \) is its envelope mass, and \( R \) its radius. The binding energy of the envelope is composed of the gravitational binding energy \( E_{\mathrm{grav}} \), thermal (internal) energy \( E_{\mathrm{th}} \), and recombination (ionization) potential energy \( E_{\mathrm{rec}} \), both modulated by their contribution fractions \( f_{\mathrm{th}} \) and \( f_{\mathrm{ion}} \). These are used to parameterise \( \lambda  \), which describes both the characteristic evolution of the binding energy and contribution from additional energy sources. These additional energy sources decrease the amount of orbital energy required for the ejection process, and thus lead to wider predicted separations.

Where computed with the \( \alpha  \)-formalism, we use the \(\lambda\) values from \citet[][see their Appendix A\footnote{The source of these values is often misreferenced as the models of \citet{dewiEnergyEquationEfficiency2000}. These \( \lambda  \) values are originally from an updated version of \texttt{BSE} used after \citet{hurleyEvolutionBinaryStars2002} was published. They are fits by Onno Pols to the same evolutionary models used for the \texttt{SSE} analytic equations (\citealt{polsStellarEvolutionModels1998,hurleyComprehensiveAnalyticFormulae2000}; O. R. Pols 2026, private communication). These \( \lambda \) values were first used in \citet{izzardNucleosynthesisBinaryStars2004} and \citet{kielPopulatingGalaxyLowmass2006}.}]{claeysTheoreticalUncertaintiesType2014} for our stars, and default to \( \lambda  = 0.5 \) for all helium stars. These assume full contribution from thermal (internal) energy sources (\( f_{\mathrm{th}} =  1.0 \)), but allow for an adjustable recombination (ionization) contribution. We take the full contribution (\( f_{\mathrm{ion}} = 1.0 \)) in our models unless stated otherwise. We found that other parametrizations for \( \lambda  \) \citep[e.g.,][]{xuBindingEnergyParameter2010,wangBindingEnergyParameter2016} predict the majority of the observed PCEB population should consist of HeWD systems at all orbital periods, which is not observed.

For the Two-stage formalism, we developed \( \lambda  \) values for the binding energy of the envelope ejected in stage I using models from the Monash Stellar Evolution code and \mesa{} for masses \( 2.25\text{--}15\,M_{\odot} \). These values are presented alongside our investigation of where the Two-stage formalism applies in Appendix \ref{sec:twostage_apply}.

\subsubsection{Critical mass ratios}
\label{sec:qcrit}
Critical mass ratios are used in rapid population synthesis to determine whether a system will enter common envelope at Roche lobe overflow \citep{hurleyEvolutionBinaryStars2002,claeysTheoreticalUncertaintiesType2014}. If the mass ratio of a system at the onset of Roche lobe overflow \( M_{\mathrm{donor}} / M_{\mathrm{accretor}} \) exceeds \(q_{\mathrm{crit}}\), the mass transfer event is expected to rapidly develop into a common envelope event \citep{hjellmingRapidMassTransfer1989}. This instability is typically evaluated from a stellar model's adiabatic response to mass loss, which recent work using detailed stellar models suggests is more stable than earlier works found using polytropic models \citep[e.g.,][]{sobermanStabilityCriteriaMass1997,chenMassTransferGiant2008,geAdiabaticMassLoss2020,temminkCopingLossStability2023}. However, using these updated criteria can produce populations which are inconsistent with observations \citep[see Sec. \ref{sec:vary_physics} and][]{yamaguchiPopulationDemographicsWhite2025}.

Developing a common envelope inspiral depends on the preceding phase of mass transfer prior to the common envelope event.
It is thus likely that overflow through the second Lagrange point (\( L_{2} \)) or non-conservative mass transfer could initiate a common envelope episode, even if dynamical instability of the mass transfer does not occur \citep{pavlovskiiMassTransferGiant2015,geThermalEquilibriumMassloss2020,marchantRoleMassTransfer2021,temminkCopingLossStability2023,hennecoContactTracingBinary2024,schurmannExploringBoundaryStable2024}. While this process can be stable overflow \citep[e.g., SS 433,][]{bowlerInterpretationObservationsCircumbinary2010}, if this mass loss is sufficiently rapid it could destabilize the system \citep[e.g.,][]{nibbsWeakenedInspiralsHigh2026}.
In the largest and most diffuse giant stars, the thermal and orbital timescales can become comparable \citep[e.g.,][]{geThermalEquilibriumMassloss2020}. Thus, if a giant star undergoes \( L_{2} \) overflow and mass loss proceeds on or near a thermal timescale, it may rapidly destabilize.
We therefore assume that common envelope occurs in giant donors which undergo \( L_{2} \) overflow, i.e.,
\begin{equation}
	q_{\mathrm{crit}} = \begin{cases}
		q_{\mathrm{ad}},                  & \text{non-giant donors}, \\
		\min(q_{\mathrm{ad}}, q_{L_{2}}), & \text{giant donors},
	\end{cases}
\end{equation}
for critical mass ratios for unstable adiabatic mass transfer (\( q_{\mathrm{ad}} \)) and \( L_{2} \) overflow (\( q_{L_{2}} \)). We take these boundaries from the Yunnan group models \citep{geAdiabaticMassLoss2015,geAdiabaticMassLoss2020,geThermalEquilibriumMassloss2020,zhangAdiabaticMassLoss2024,zhangErratumAdiabaticMass2026} and use values from \citet{temminkCopingLossStability2023} where the Yunnan grids do not cover in \( \log R\), which only occurs at the tip of the giant branches in \( M_{1,0} \lesssim 1.1 M_{\odot} \) binaries. We prefer the Yunnan group values from \added{fully conservative mass loss sequences} for consistency \added{and completeness} across all stellar types, \added{though the stability can vary with the mass transfer efficiency \citep[e.g.,][]{hjellmingThresholdsRapidMass1987,geAdiabaticMassLoss2024}}.
We test adjustments to this assumption in Sec. \ref{sec:vary_physics}.

\subsection{Forward-modelling the observed population}
\begin{figure*}[t]
	\centering
	\includegraphics[width=\textwidth]{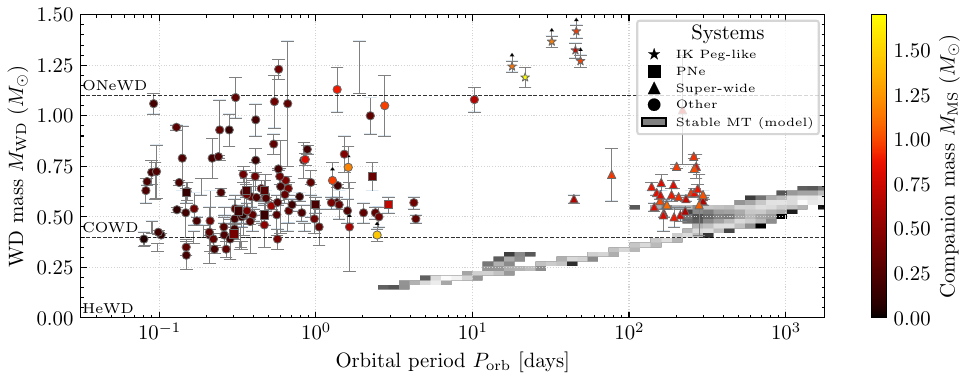}
	\caption{Points are compiled observations of WD-MS post-common envelope binaries (PCEBs) \citep{zorotovicPostcommonenvelopeBinariesSDSS2010,nebotgomez-moranPostCommonEnvelope2011,rebassa-mansergasMagnitudelimitedCatalogueUnresolved2025,jonesObservationalConstraintsCommon2020,hernandezWhiteDwarfBinary2021,hernandezWhiteDwarfBinary2022a,yamaguchiWidePostcommonEnvelope2024,yamaguchiWidePostcommonEnvelope2024a,booneSearchingGEMSDiscovery2026,shariatGlobalViewPostinteraction2026,shiraishiTwoUnseenMassive2026,motherwayNotsocompactCompanionMassive2026}. Notable PCEB types are denoted by specific markers\added{: IK Peg-like systems (stars), planetary nebulae (squares), and super-wide orbits (triangles)}. If a companion mass was not estimated, the point is given by a white marker. \added{A 2D histogram of stable mass transfer products from our models for masses \( 0.8 < M_{\mathrm{MS}} < 1.2M_{\odot} \) is also shown (grey rectangles).} Some of the PCEBs wider than \( 10^{2}  \) days could be formed through both common envelope and stable mass transfer\added{, as they overlap part of the predicted stable mass transfer population.}}
	\label{fig:observations}
\end{figure*}
\begin{figure*}[t]
	\centering
	\includegraphics[width=\textwidth]{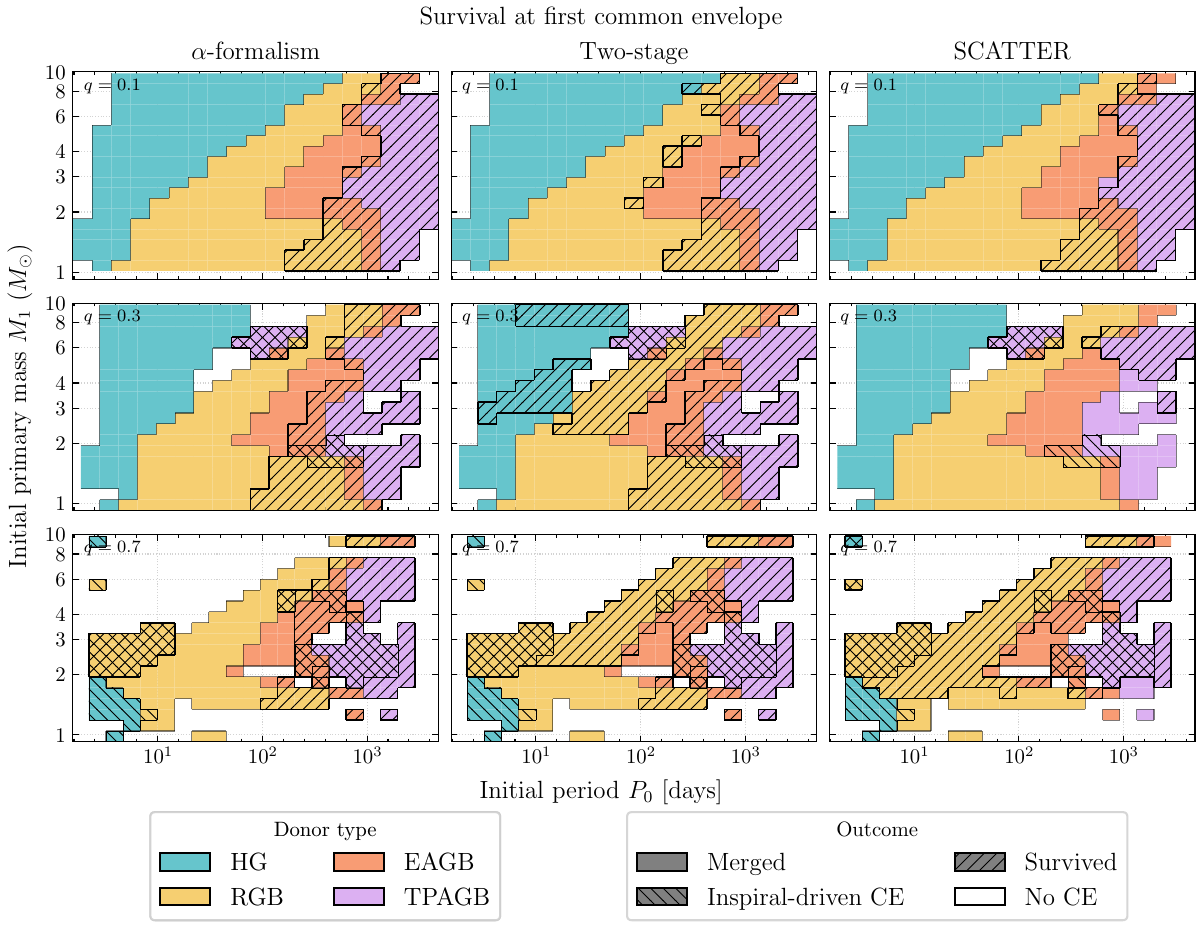}
	\caption{\added{Survival of the first common envelope event for three different mass ratios \( q \in \{0.1,0.3,0.7\}   \) (rows) across the initial binary parameter space, colored by the donor's stellar type.} Regions are hatched if the binary survives common envelope, or if the common envelope event was initiated by inspiral from a continued episode of non-conservative stable mass transfer. The Two-stage formalism predicts more systems survive HG and RGB common envelope episodes, and the SCATTER formalism predicts that typically only AGB common envelope episodes are survived.}
	\label{fig:survival}
\end{figure*}
The \binc{} results are generalized to a physical population by weighting each simulated binary to their natal (birth) probabilities \citep{kempImpactMetallicityNova2022,osbornUsingBinaryPopulation2025}. For each system \( j \), this is a weight per star forming mass, given by
\begin{equation}
	w _{j} = \omega_{\mathrm{m}}\,\frac{f_{b}}{N} \cdot \frac{\pi (\mathbf{x}_{j})}{\zeta (\mathbf{x}_{j})}
	\label{eq:weight}
	,\end{equation}
where \( \pi(\mathbf{x}_{j}) \) and \( \zeta(\mathbf{x}_{j}) \) are the natal and sampling probabilities of parameters \( \mathbf{x}_{j} = \{M_{1,0},q_{0},\log P_{0}\}   \) (see Table \ref{tab:params}), \( N \) is the total number of systems, \( f_{b} \) is the binary fraction of the population, and \( \omega_{\mathrm{m}} \) is the mass normalization term that describes the number of stellar systems forming per \( M_{\odot} \) of star-forming material. In all weightings, we assume the binary fraction as 50\% \citep{offnerOriginEvolutionMultiple2023}. Using empirical weights for \( f_{b} \), \( q_{0} \) and \( P_{0} \) \citep[e.g.,][]{moeMindYourPs2017} did not change our conclusions.

Present-day populations are then created by convolving the weights across a given star formation history.
In other words, at a given time \( t_{n} \) over a given binning in parameters \( \boldsymbol{\theta} \), we look at the population formed at all preceding bins \( k \in \{1,\ldots,n\}\) at time \( t_{k} \), and see what those stars are at \(t_{n} \), which is given by the sum of weights \( w_{j} \) in that bin at \( t_{n} \).

\subsubsection{Solar neighbourhood model}
\label{sec:sfr}
The majority of observed post-common envelope binaries lie within 1 kpc of the Sun. As such, we consider the stellar density within the \emph{`Solar neighbourhood'} --  a cylindrical volume centered on the Sun out to 1 kpc in both height and radius. We assume stellar density declines exponentially above and below the Galactic plane symmetrically with height \( z \) \citep[e.g.,][]{paczynskiTestGalacticOrigin1990,toonenEffectCommonenvelopeEvolution2013,neijsselEffectMetallicityspecificStar2019,songBinaryPopulationSynthesis2025},
\begin{equation}
	\rho_{\star }(z) = \rho_{0} \exp \left( - \frac{|z|}{h_{z}} \right)
	,\end{equation}
where we set the scale height \( h_{z} = 380\,\mathrm{pc} \) \citep[e.g.,][]{mckeeStarsGasDark2015,mackerethAgeMetallicityStructure2017}. Assuming the midplanar stellar density is \( \rho_{0} = 0.043 \,M_{\odot}\,\mathrm{pc^{-3} }\) as computed for this scale height \citep{mckeeStarsGasDark2015}, the total present-day mass in our 1 kpc cylinder is \( 9.57\times 10^{7}M_{\odot}  \). This is in line with Galactic estimates of the total thin disc mass, as our considered volume is roughly 1/100th its size \citep{bland-hawthornGalaxyContextStructural2016}.

The star formation rate is taken per the Galactic chemical evolution model data of \citet{kobayashiOriginElementsCarbon2020}. The star formation rate is an output of their model, which can reproduce observed chemical abundances and metallicity distributions in the Solar neighbourhood\footnote{The model data is available at \href{https://star.herts.ac.uk/~chiaki/gce/}{star.herts.ac.uk/\textasciitilde{}chiaki/gce/}. For more details about this Solar neighbourhood model, see Sec. 2.2 of \citet{kobayashiOriginElementsCarbon2020} and references therein.}. We normalize the rate such that the total formed mass is equal to our chosen mass of \( 9.57\times 10^{7}\,M_{\odot}  \) for the Solar neighbourhood at 13.5 Gyr.

\subsubsection{The magnitude-limited population}
We adjust each system's weight based on how magnitude-limited our defined Solar neighbourhood is for that system type. This weight represents how observable the sub-population represented by system \( j \) is with current astronomical instrumentation. We define the weight as the ratio of the surveyable to total stellar volume for each system's absolute magnitude \( M \), which determines a maximum surveyable distance \( d_{\mathrm{max}} \).
\begin{equation}
	w_{\mathrm{lim},j}(M) =  \min \left[\frac{\int_{V_{\mathrm{sph}}(d_{\mathrm{max}})} \rho_{\star }(z) \,\mathrm{d} V_{\mathrm{sph}} }{\int_{V_{\mathrm{sol}}} \rho_{\star }(z)  \,\mathrm{d} V_{\mathrm{cyl}}}, 1 \right]
	\label{eq:limit_mag}
	,\end{equation}
where \( \mathrm{d}V_{\mathrm{sph}} \) and \( \mathrm{d}V_{\mathrm{cyl}} \) are the cylindrical and spherical volume elements, \( V_{\mathrm{sol}} \) is the complete volume of our 1 kpc Solar neighbourhood, and \( V_{\mathrm{sph}}(d_{\mathrm{max}}) \) defines the magnitude-limited footprint as a spherical volume of radius \( d_{\mathrm{max}}(M) \). The ratio is independent of the assumed \( \rho_{0} \). We take Gaia \( G = 20 \) as our limiting magnitude \citep{rielloGaiaEarlyData2021}.

We do not forward model a ``distinguishability" criterion based on the UV excess as done in previous studies \citep[e.g.,][]{davisComprehensivePopulationSynthesis2010,toonenEffectCommonenvelopeEvolution2013}. Some observed systems are discovered serendipitously without systematic searches \citep[e.g.,][]{wonnacottIKPegNearby1993,shiraishiTwoUnseenMassive2026} or did not require UV photometry to flag it as a candidate post-common envelope binary \citep[][see \citealt{shahafTriageAstrometricBinaries2019}]{yamaguchiWidePostcommonEnvelope2024}. The selection function is complex and varies across the observable parameter space, but remains an opportunity for future investigation.
Given we do not model the selection effects from both distinguishability and survey cadence, we expect our model to predict more systems than what is observed if the physics is correct.

\section{Results}
\label{sec:results}
\subsection{Observational compilation of WD-MS binaries and the stable mass transfer boundary}
\label{sec:obs}
We have compiled a series of \npcebtot{} observed, detached PCEBs with well-characterized masses to compare to our present-day model, shown in Fig \ref{fig:observations}. We restrict ourselves to detached, nearly circular binaries which are most likely to be from common envelope and have not interacted significantly since the event. Our compilation includes the well-studied sample of post-common envelope binaries \citep{zorotovicPostcommonenvelopeBinariesSDSS2010}, the 58 SDSS post-common envelope binaries selected via photometry and RV variation \citep{nebotgomez-moranPostCommonEnvelope2011}, 57 new binaries visually selected from the \emph{Gaia} bands with eclipsing light-curves from the Zwicky Transient Facility (ZTF) \citep{rebassa-mansergasMagnitudelimitedCatalogueUnresolved2025,shariatGlobalViewPostinteraction2026}\footnote{We removed Gaia DR3 2273583445431091584, as it appears to be cyclotron-emitting and thus likely semi-detached.}, nine central binaries of planetary nebulae with well-constrained masses \citep{jonesObservationalConstraintsCommon2020}\footnote{We exclude M3-1 and NGC 6337 as these stars were analyzed assuming a constant primary mass}, the 5 ONeWD systems from \citet{yamaguchiWidePostcommonEnvelope2024a}, the 31 new wide \emph{Gaia} binaries with G-type companions \citep{yamaguchiWidePostcommonEnvelope2024}, and an additional nine binaries from other works \citep{hernandezWhiteDwarfBinary2021,hernandezWhiteDwarfBinary2022a,booneSearchingGEMSDiscovery2026,shiraishiTwoUnseenMassive2026,motherwayNotsocompactCompanionMassive2026}.

We denote systems of certain unique types. Those with both massive companions (G-type or earlier, \( M_{\mathrm{MS}} \gtrsim 0.8M_{\odot} \)) and massive WD's are termed ``IK Peg-like", in reference to IK Peg \citep{wonnacottIKPegNearby1993}. Observations of these systems originally motivated additional energy sources for the ejection of the common envelope \citep[e.g.,][]{davisComprehensivePopulationSynthesis2010,zorotovicPostcommonenvelopeBinariesSDSS2010,demarcoFormalismCommonEnvelope2011}. The few planetary nebulae (PNe) systems from \citet{jonesObservationalConstraintsCommon2020} are highlighted but show no notable differences with similar non-PNe PCEBs. The majority of observed systems are COWD systems, and only five have primary masses consistent with a HeWD.

Paradoxically, the most observed populations are on opposite sides of the \( P_{orb} \) space -- those found from systematic radial velocity surveys which are most sensitive to the sub-day, cataclysmic variable-like orbits (`CV-like', \( P < 1 \) days), and those found through Gaia astrometry (\emph{`super-wide'}, \( P > 100 \) days). These samples are likely the most observationally complete, with incompleteness increasing to higher companion masses and the intermediate period range of roughly 3 to 100 days.

We also plot non-CE stable mass transfer systems from our model in gray bins, which overlay some of the super-wide PCEBs with G-type companions from \citet{yamaguchiWidePostcommonEnvelope2024}. This suggests they may arise from both stable mass transfer and common envelope. The stable mass transfer channel may involve moderate eccentricities as opposed to nearly circular systems, given a natal velocity boost from asymmetric mass loss at the end of the AGB if mass transfer truncates early \citep[e.g.,][]{el-badryEmpiricalMeasurementInitialFinal2018,chawlaGaiasPromiseDetect2026,oconnorFateGaiasWide2026} or interactions with a circumbinary disc as observed in post-AGB binaries \citep{vanwinckelPostAGBBinariesInteracting2025}. \added{Also, some of the binaries may form through pathways involving mergers in hierarchical triples \citep[e.g.,][]{hallakounIKPegasiDoublemerger2026}, but we aim to test which PCEBs can be formed from isolated binaries under each formalism.}

\subsection{Changes in common envelope survival and progenitor pathways}
\label{sec:survival_pathways}
\begin{figure*}[t]
	\centering
	\includegraphics[width=\textwidth]{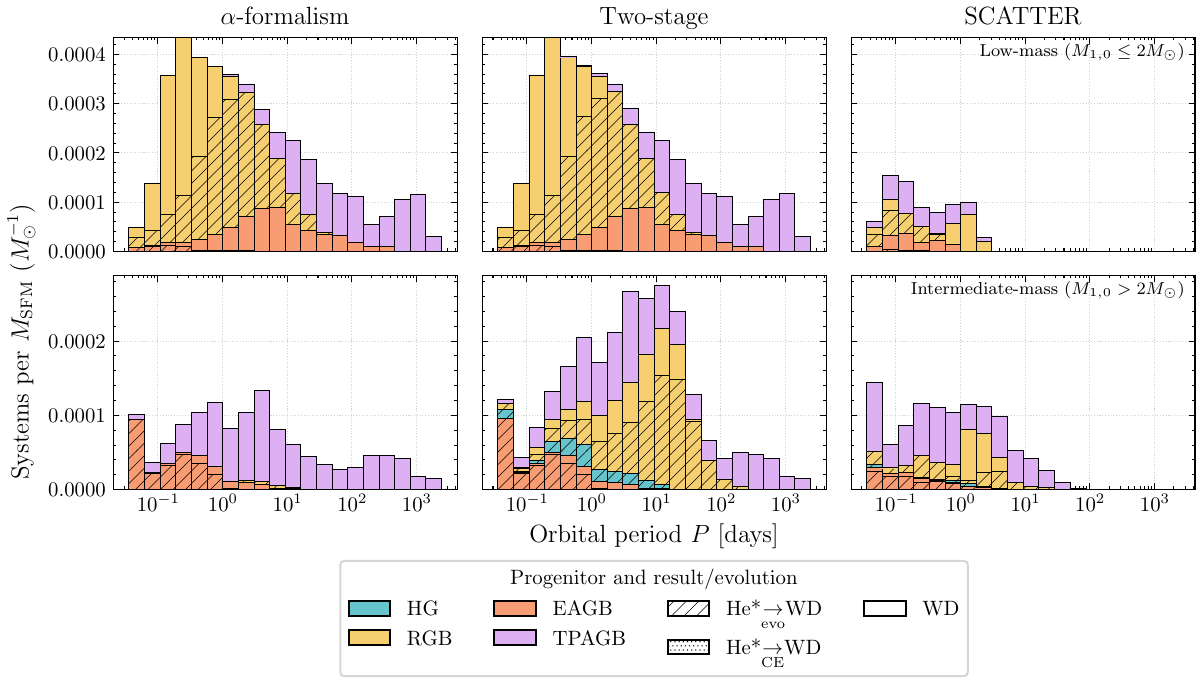}
	\caption{\added{Evolutionary pathways for present-day PCEBs in the orbital period marginal, for low-mass (top panels) and intermediate-mass (bottom panels) donors.} The Two-stage formalism increases how many systems with intermediate-mass primaries survive first common envelope in close orbits (as per Fig. \ref{fig:survival}), and the SCATTER formalism predicts the majority of PCEBs should be from TPAGB donors but never in wider orbits than \( 10^{2}\) days. Some pathways involve either a second common envelope or the static evolution of a naked helium burning star to a white dwarf.}
	\label{fig:pathways}
\end{figure*}
We first assess how each formalism changes which systems survive common envelope as detached binaries and their progenitor pathways. In Fig. \ref{fig:survival}, we show which systems in our parameter space survive their first common envelope, where the color represents the donor's stellar type at the onset of unstable mass transfer.

For the Two-stage formalism, common envelope during the HG and RGB phases becomes survivable toward higher masses across all mass ratios, whereas it is not under the standard \( \alpha  \)-formalism. A clear boundary is visible where we apply the Two-stage formalism to stars more massive than \( 2.25 M_{\odot} \).
The SCATTER formalism only predicts that common envelope is rarely survived except in AGB donors, typically only survived in other donors for higher mass ratios (\( q \gtrsim 0.6 \)).

There are also regions of parameter space where the initial episode of mass transfer is stable, but a common envelope event eventually occurs as the systems inspiral from the orbital angular momentum loss of non-conservative mass transfer. We refer to these as \emph{inspiral-driven common envelopes}, which will always enter their first common envelope with a reduced envelope mass. Typically, around 45\% of the envelope mass can be lost in this pathway, with at most 70\%. Survivors of this pathway are visible across intermediate-mass (\( M_{1,0} \sim 7\,M_{\odot} \)) TPAGB donors at \( q=0.3 \) and low-mass (\( M_{1,0} \sim 2.5\,M_{\odot} \)) RGB and TPAGB donors at \( q =0.7 \).
\begin{figure*}[t]
	\centering
	\includegraphics[width=\textwidth]{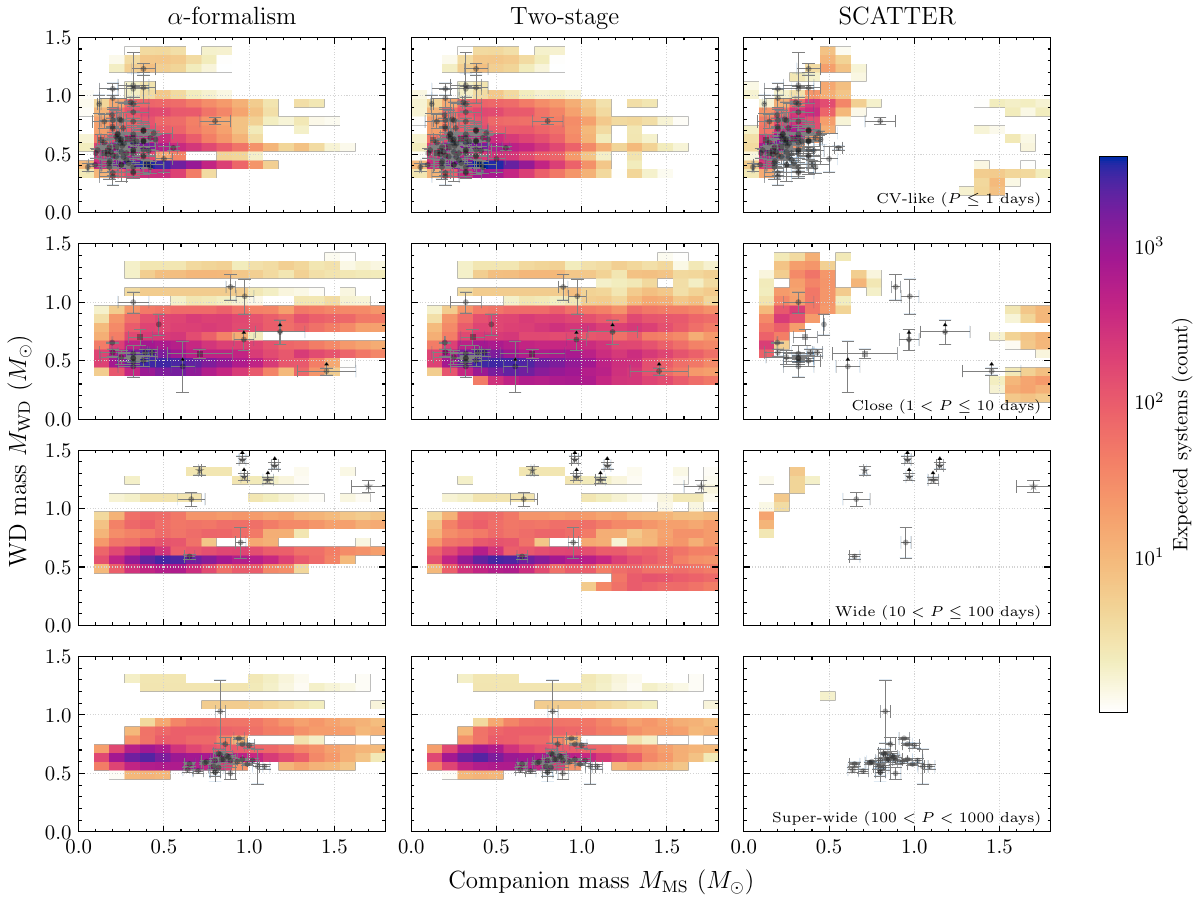}
	\caption{\added{Present-day population of PCEBs in observed mass space over four different ranges of orbital period \( P \) (rows) for each formalism (columns)}. The observed PCEB population favours the \( \alpha  \)-formalism or the Two-stage formalism, as the SCATTER formalism (using Eq. \ref{eq:etafit}) does not predict overdensities where observed. Markers for observed systems are per Fig. \ref{fig:observations}.}
	\label{fig:presentday_mass}
\end{figure*}

These changes in common envelope survival alter the possible progenitor pathways for PCEBs, so we show the proportions of progenitor pathways for detached PCEBs across their mean orbital period \( P_{\mathrm{PCEB}} \) as a PCEB system in Fig. \ref{fig:pathways}. Some pathways involve either a second common envelope or the static evolution of a naked helium burning star to a white dwarf.

The distributions for low-mass stars (\( M \le 2\, M_{\odot} \)) are identical for both the \( \alpha  \)-formalism and Two-stage formalism, which is expected since they are computed identically. Low mass stars favour mostly RGB progenitors, as they expand to much greater radii than their intermediate-mass RGB counterparts. The majority of close PCEBs are formed from low-mass primaries, which undergo a common envelope episode from an RGB to a WD (\( \text{RGB}\!\underset{\text{CE}}{\rightarrow}\!\text{WD}\)), which form the bulk of HeWD PCEBs in CV-like orbits (\( P \lesssim \) 2 days). Those that form helium stars (\( \text{RGB}\!\underset{\text{CE}}{\rightarrow}\!\text{He*} \)) are observed as sdOBA-type hot subdwarf stars, which evolve in COWDs quiescently (\( \text{He*}\!\underset{\text{evo}}{\rightarrow}\!\text{WD} \)). Roughly 20\% of these later evolve into double degenerate COWD/HeWD systems \citep[e.g.,][]{andrewsMassDistributionCompanions2014,brownELMSurveyVII2016a,el-badryBirthELMsZTF2021}.

With the energy-based formalisms, the super-wide PCEBs, with \( P > 100 \), days solely arise from TPAGB donors. This is consistent with previous works \citep[e.g.,][]{davisComprehensivePopulationSynthesis2010,toonenEffectCommonenvelopeEvolution2013,yamaguchiWidePostcommonEnvelope2024,belloniFormationLongperiodPostcommon2024}. These systems are expected out to separations of over 1,000 days.
However, the SCATTER formalism does not predict any PCEBs with orbital periods longer than \( 100  \) days. This suggests the observed super-wide PCEBs with periods longer than 100 days \citep[e.g.,][]{yamaguchiWidePostcommonEnvelope2024} cannot be explained by the SCATTER formalism.

With both the \( \alpha  \) and Two-stage formalism roughly one-third of binaries between 1 and 100 days result in naked helium-burning stars (i.e., \( 0.45 < M < 1.2\,M_{\odot}\)), shown as hatched regions (\( \text{RGB}\!\underset{\text{CE}}{\rightarrow}\!\text{He*} \)). These are observable for a brief window of \( \sim\!1\text{--}100 \) Myr \citep{woosleyEvolutionMassiveHelium2019}. In extremely rare cases under the Two-stage formalism, the naked helium star is massive enough to both undergo a giant phase and a second CE, making it possible for a binary to undergo two common envelope episodes on the same star.

The Two-stage formalism mainly increases the number of intermediate-mass systems which survive in close orbits, as shown in Fig. \ref{fig:survival}. Since HG and RGB common envelope events are survivable for intermediate-mass donors (Fig. \ref{fig:survival}), there are now over double the number of wider PCEBs between 1 and 100 days  from these surviving intermediate-mass donors, which evolve to white dwarfs quiescently (\( \text{He*}\!\underset{\text{evo}}{\rightarrow}\!\text{WD} \)). These arise from the stage II mass transfer present in intermediate-mass donors (see Sec. \ref{sec:intro_twostage}), which leads to an increase in the separation after the convective envelope's ejection (i.e., \( a_{\mathrm{f}} / a_{\mathrm{s_{1}}} > 1 \) in Eq. \ref{eq:twostage}) \citep{hiraiTwostageFormalismCommonenvelope2022,pickerFitsConvectiveEnvelope2024}.
We expect these to maintain a hydrogen-rich atmosphere and appear as a DA-type white dwarf \citep[e.g.,][]{millerbertolamiPrimerFormationEvolution2024}, and thus be observationally indistinguishable from other pathways.
Survival of these systems in high \( q \) binaries could increase the number of Type Ia supernovae via the single degenerate channel \citep{ruiterDelayTimesRates2011,claeysTheoreticalUncertaintiesType2014}.

\subsection{The parameter distributions of present-day PCEBs}
\label{sec:paramdist}
\begin{figure*}[htpb]
	\centering
	\includegraphics[width=\textwidth]{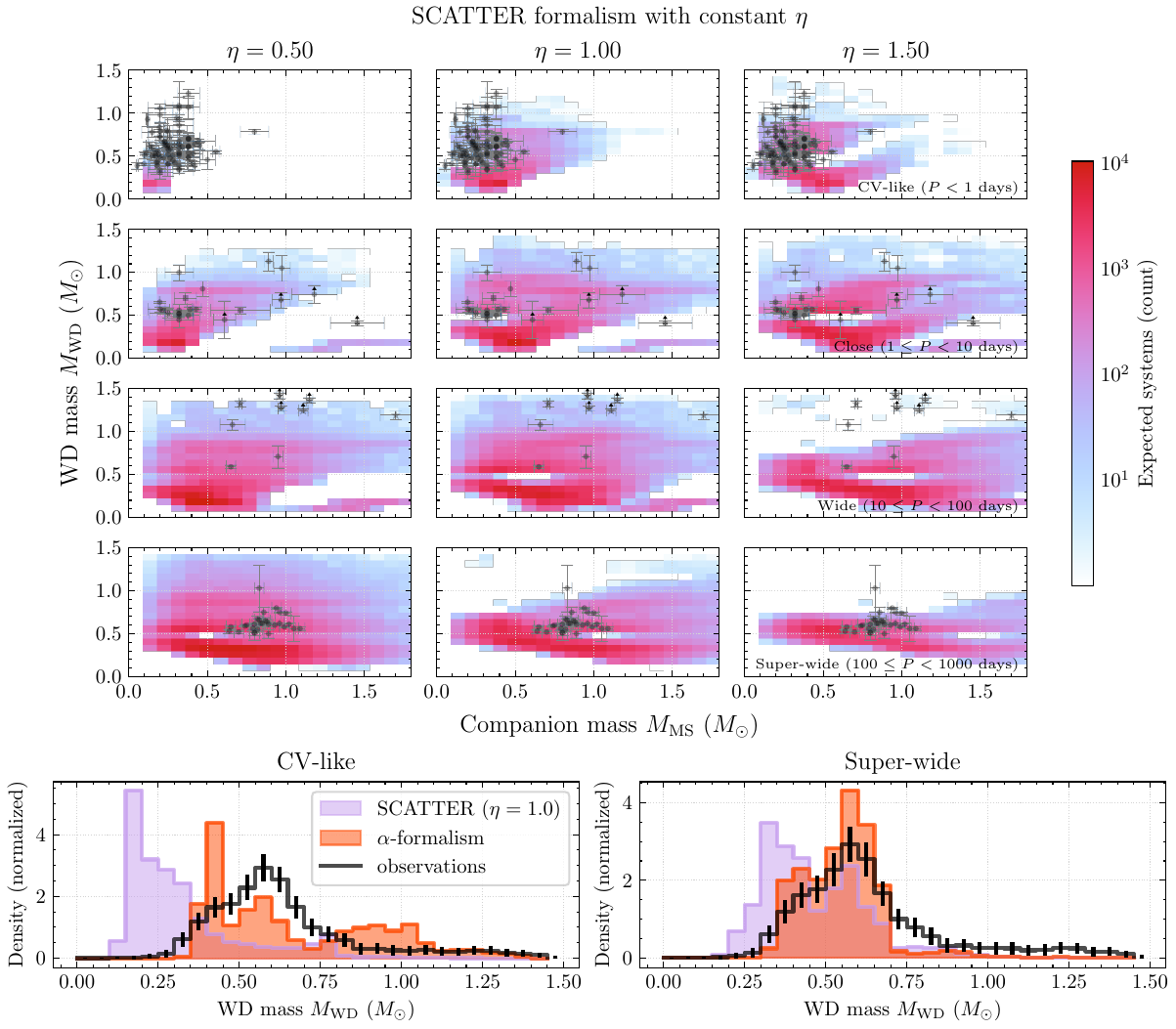}
	\caption{\added{The SCATTER formalism, with different values for \( \eta  \) (columns, see Eq. \ref{eq:scatter}), instead of the fitted \( \eta  \) (Eq. \ref{eq:etafit}) used in Fig. \ref{fig:presentday_mass}.} While the observed population \emph{appears} well-matched in observed mass space by a constant \( \eta \approx 1.0  \), gaps are present at the bulk of the observed close and CV-like PCEBs (top panels). The model also predicts the less massive HeWD primaries outnumber COWDs at all orbital periods (bottom panels). HeWDs are predicted to comprise 45\% of all PCEBs and 76.7\% of CV-like PCEBs at \( \eta =1.0 \), which is not reflected in the observed population(s). Markers for observed systems are per Fig. \ref{fig:observations}.}
	\label{fig:eta_sweep}
\end{figure*}

Fig. \ref{fig:presentday_mass} shows component mass distribution of the present-day WD-MS PCEB population given by our forward-modelling criteria in different orbital period ranges. For clarity, we refer to each order of magnitude slice in orbital period as CV-like (\( <\!1 \) day), close (\( 1\text{--}10 \) days), wide (\( 10\text{--}100 \) days), or super-wide (\( 100\text{--}1000 \) days) following the notation given in the right-most column of the Figure.

Overdensities of predicted surviving PCEBs in models with energy-based formalisms align well with the observed samples at CV-like and close periods for M-dwarfs, which is the most complete parameter space for observations \citep{shariatGlobalViewPostinteraction2026}. Our standard model uses \( \alpha_{\mathrm{CE}} = 0.2 \) from the constraints from previous studies \citep[e.g.,][]{zorotovicPostcommonenvelopeBinariesSDSS2010,demarcoFormalismCommonEnvelope2011}, which appears to align well for these systems.
%
Many PCEBs are predicted where observations are absent, across wide PCEBs (10-100 days) and/or companion masses above \( 0.5 M_{\odot} \). Systematic surveys over these regions are challenging and likely incomplete, as they require long-baseline observations on unresolved systems where a MS companion dominates the flux.
The Two-stage formalism predicts larger overdensities in these regions, mainly for FG-type companions (\( M_{\mathrm{MS}} > 0.8\,M_{\odot} \)) at close and wide orbital periods, as stated in Sec. \ref{sec:survival_pathways} (see Figs. \ref{fig:survival} \& \ref{fig:pathways}). Since there are only two (lower-limit) observations in this mass/period range, there is no strong observational support in WD-MS PCEBs for the Two-stage formalism.
The SCATTER formalism overdensities clearly do not align with observations.

At super-wide orbits, both energy-based formalisms predict PCEBs with G-type and earlier companions, whereas the SCATTER formalism does not predict (almost) any.
PCEBs are modelled where observed, but predictions also extend to lower and higher companion masses.
Those with lower companion masses may be excluded from the observational selection as they cannot be confidently distinguished from hierarchical triples with an inner MS binary \citep[see][]{shahafTriageGaiaDR32024,yamaguchiWidePostcommonEnvelope2024},
and those with early spectral types are beyond the \( 1.2 M_{\odot} \) upper limit recoverable from the \emph{Gaia} astrometric flux excess \citep{shahafTriageAstrometricBinaries2019}.

Our model struggles to replicate the IK Peg-like systems (\( M_{\mathrm{WD}} \gtrsim 1.1 M_{\odot}, M_{\mathrm{MS}} \sim 1.4 M_{\odot} \)) in wide orbits (20--80 days) in any formalism. We discuss this further in Sec. \ref{sec:ikpeg}.

\section{Discussion}
\label{sec:discussion}
\subsection{Where the angular momentum budget fails}
\label{sec:angmom}

Our results and previous works \citep[e.g.,][]{nelemansReconstructingEvolutionWhite2005,webbinkCommonEnvelopeEvolution2008,davisComprehensivePopulationSynthesis2010,zorotovicPostcommonenvelopeBinariesSDSS2010,woodsCanWeTrust2011,ruiterDelayTimesRates2011,ivanovaCommonEnvelopeEvolution2013,toonenEffectCommonenvelopeEvolution2013} suggest that orbital angular momentum-based formalisms struggle to describe common envelope outcomes. In our model, the SCATTER formalism cannot explain the existence of hot subdwarfs in close binaries and extremely low-mass white dwarfs (Fig. \ref{fig:survival}), nor wide and high mass white dwarf systems (Fig. \ref{fig:presentday_mass}).

\begin{figure*}[htpb]
	\centering
	\includegraphics[width=\textwidth]{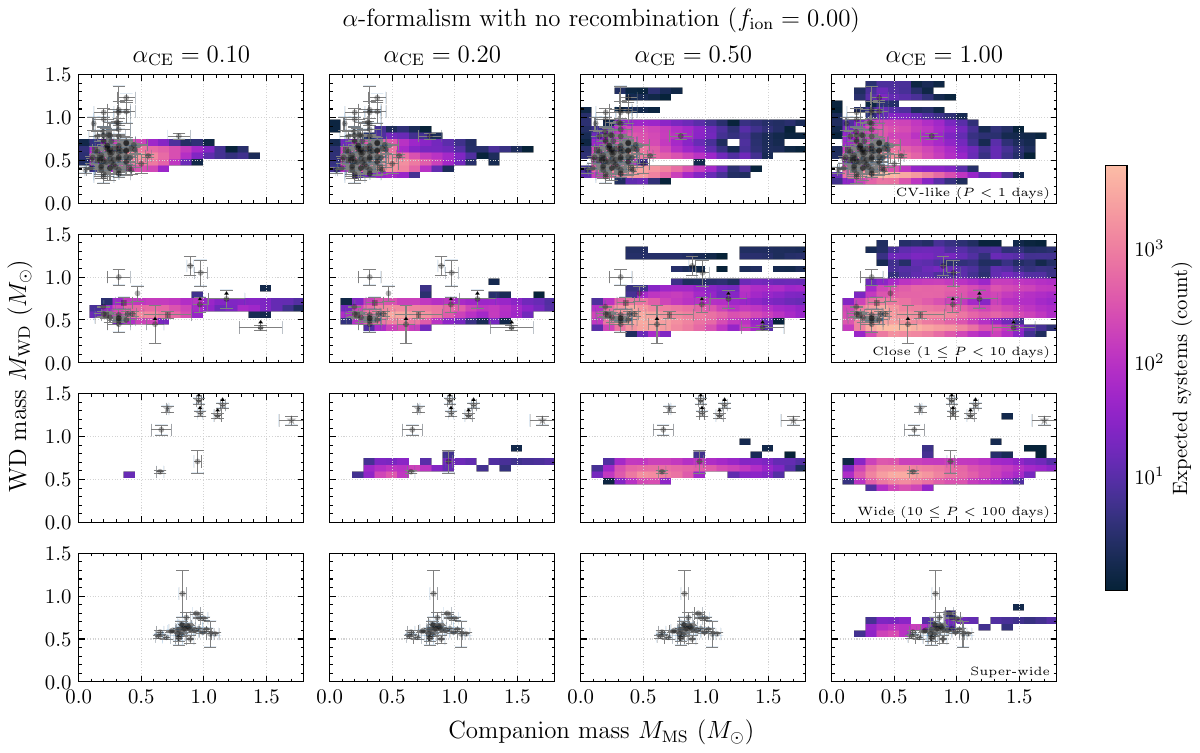}
	\caption{Recombination energy is necessary to reproduce the observed population of PCEBs -- no model using the \( \alpha  \)-formalism without recombination energy reproduces the wide IK Peg-like and all super-wide PCEBs over any  \( 0 < \alpha_{\mathrm{CE}} < 1  \) (columns). Markers for observed systems are per Fig. \ref{fig:observations}.}
	\label{fig:claeys_frec_0}
\end{figure*}

The functional \( \eta  \) fit we use in our standard model could be miscalibrated. Indeed, we find that the functional \( \eta  \) predicts the orbit should shrink by a factor of 1000 when \( q_{\mathrm{c c}} = M_{\mathrm{core}}/M_{2} \sim 1.0 \), which explains why no RGB donors with a radius of about 100 \( R_{\odot} \) survive common envelope where \( q_{0} \) is similar to the core mass (i.e., \( q\sim 0.4 / M_{1} \) for the typical low-mass RGB core) (Fig. \ref{fig:survival}).
We test constant values for \( \eta  \) in Fig. \ref{fig:eta_sweep}, but we find that no value for \( \eta \) reproduces all the observed overdensities. Higher values of \( \eta > 1.5 \) predict no high mass WDs survive in the regions of mass space where observed, and lower values do not predict close PCEBs with massive white dwarfs.

The observed population is best matched by a constant \( \eta \simeq 1.0 \), though with gaps predicted where the bulk of CV-like PCEBs are observed. While the model \emph{appears} well-matched apart from this region across the mass space, the predicted proportion of HeWD to COWD primaries is substantially different from observations.
Across all periods, the largest overdensities are around low-mass HeWDs from low-mass HG/RGB donors instead of those for COWDs, shown in the WD mass distribution marginals (bottom panels, Fig. \ref{fig:eta_sweep}). HeWDs comprise \( \sim\! 46\% \) of all surviving PCEBs across all orbital periods, and \( 76.7\% \) of the CV-like PCEB population. To our knowledge, there is no significant population of these HeWD+dM systems in the literature at any orbital period \citep[e.g.,][]{rebassa-mansergasPostcommonEnvelopeBinaries2011,kupferHotSubdwarfBinaries2015,heberHotSubluminousStars2016,rebassa-mansergasMagnitudelimitedCatalogueUnresolved2025,shariatGlobalViewPostinteraction2026}, and current observations of PCEBs do not show these extreme proportions. As the \( \alpha\)-formalism models the proportions and mass-space more closely to those observed, we do not favour the SCATTER formalism over the energy-based formalisms.

These results, along with previous ones for the \( \gamma \) formalism \citep[e.g.,][]{davisComprehensivePopulationSynthesis2010,ruiterDelayTimesRates2011,toonenEffectCommonenvelopeEvolution2013}, suggest that using solely the orbital angular momentum budget is challenging to predict common envelope outcomes. While both energy and angular momentum conservation are physically respected, it may be that the orbital angular momentum is not fully \emph{predictive} of all outcomes. Parametrizing based on the orbital angular momentum budget gives rise to exponential terms, which are challenging to constrain the parameters of from observations \citep{webbinkCommonEnvelopeEvolution2008,ivanovaCommonEnvelopeEvolution2013}.

\subsection{The widest systems and the recombination energy contribution}
\label{sec:ikpeg}
\begin{figure*}[htpb]
	\centering
	\includegraphics[width=\textwidth]{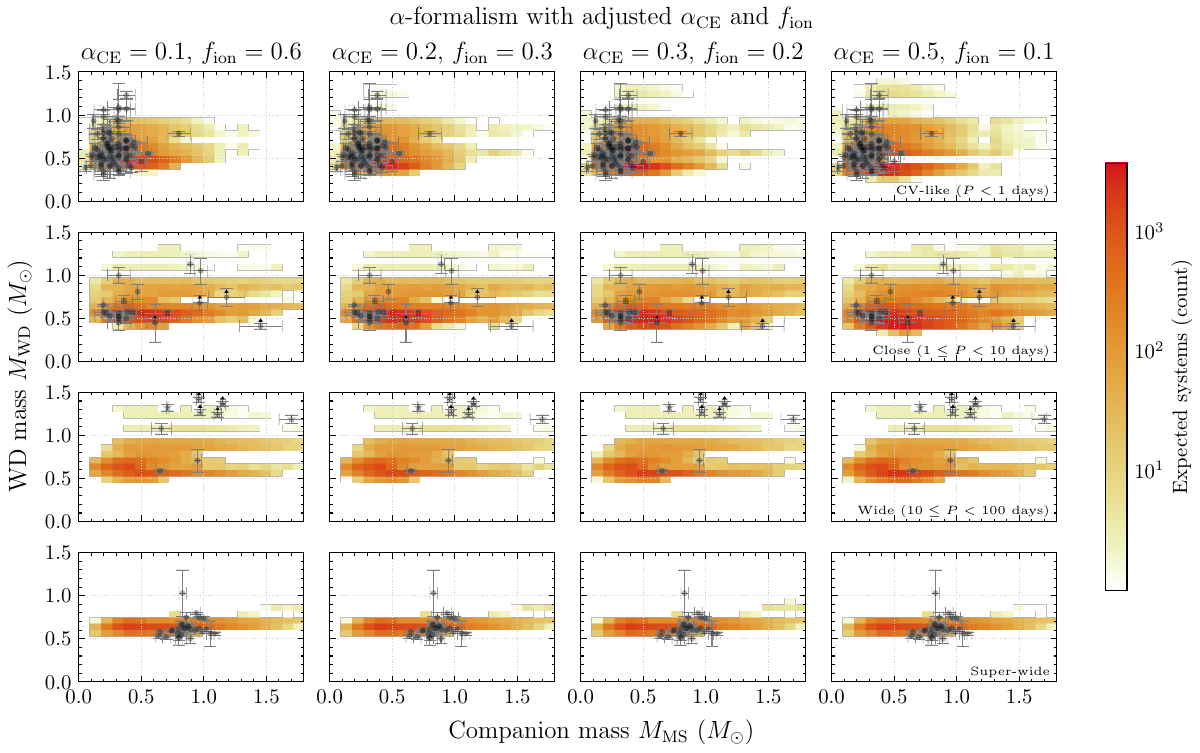}
	\caption{Four different pairs for efficiency \( \alpha_{\mathrm{CE}} \) and recombination energy contribution \( f_{\mathrm{ion}} \) (columns) which can minimally reproduce the observed PCEB population. The minimum required recombination contribution increases with decreasing values for \( \alpha_{\mathrm{CE}} \), but we do not favour \( \alpha_{\mathrm{CE}} \gtrsim 0.5 \) as it begins predicting HeWDs with K-type companions in CV-like orbits.}
	\label{fig:claeys}
\end{figure*}
Our standard model with the \( \alpha  \)-formalism does not predict the presence of most IK Peg-like systems (\( M_{\mathrm{WD}} \gtrsim 1.1 M_{\odot}, M_{\mathrm{MS}} \sim 1.4 M_{\odot}, P \sim 20\text{--}80\,\mathrm{days} \)). Unlike previous works which found progenitors of IK Peg do not survive common envelope \citep{davisComprehensivePopulationSynthesis2010,zorotovicPostcommonenvelopeBinariesSDSS2010,demarcoFormalismCommonEnvelope2011}, our inclusion of a full recombination energy contribution leads to almost all super-AGB donor systems surviving common envelope (Fig. \ref{fig:survival}). Their ONeWDs above \( M_{\mathrm{WD}} > 1.1M_{\odot} \) populate close and super-wide orbital periods (Fig. \ref{fig:presentday_mass}) instead of the wide IK Peg-like orbital periods.
This mismatch in where the ultra-massive ONeWDs appear in our model suggest that adjustments to \( \alpha_{\mathrm{CE}} \) and/or the recombination contribution \( f_{\mathrm{ion}} \) are necessary. As such, we investigate variations to both parameters.

We first assess if recombination is \emph{necessary} in Fig. \ref{fig:claeys_frec_0}.
It has been recently claimed that the widest PCEBs do not need recombination energy for their formation, though requiring a near unity efficiency \citep[e.g., \( \alpha_{\rm CE} \sim 0.9 \);][]{belloniFormationLongperiodPostcommon2024}. We cannot replicate these results -- we find that a recombination contribution is necessary to predict the full population regardless of the value of \( \alpha_{\mathrm{CE}} \). No model can completely reproduce IK Peg-like PCEBs and super-wide PCEBs without recombination at any \( \alpha_{\mathrm{CE}} \). The difference between our results likely arises from our forward-model approach: while it may be possible, it is not probable to form these systems within the considered stellar volume so we predict to see none.
We thus conclude that common envelope evolution requires \emph{some} contribution from recombination energy to form the observed PCEB population.

If recombination is necessary, how strong is its contribution to the final separation?
\( \alpha_{\mathrm{CE}} \) and \( f_{\mathrm{ion}} \) are naturally correlated per Eqs. \ref{eq:alpha} and \ref{eq:ebind}, so there are many pairs which can replicate the population.
We present representative examples of these models in Fig. \ref{fig:claeys}. The values are inversely correlated -- lower values of \( \alpha_{\mathrm{CE}} \) with higher values for \( f_{\mathrm{ion}} \) can replicate the population, and vice-versa. However, we do not favour \( \alpha_{\mathrm{CE}} \gtrsim 0.5 \) as these begin to predict large numbers of HeWD primaries in CV-like orbits (top right panel), which are not observed.
Assuming the \citet{shahafTriageGaiaDR32024} Gaia selection used to construct the super-wide sample of \citet{yamaguchiWidePostcommonEnvelope2024} is observationally complete, \( f_{\mathrm{ion}} \) can be no higher than \( \sim\!0.4 \) at \( \alpha_{\mathrm{CE}} = 0.2 \). Higher contributions predict PCEBs of IK Peg-like masses at super-wide orbital periods, but this will be necessary if such systems are observed.
A higher \( \alpha_{\mathrm{CE}} \) could also be possible if a population of CV-like PCEBs with HeWD primaries and FGK-type companions exist, but we do not favour this currently since so few are observed.
The rough minimum in both parameters required to reproduce the observed population is around \( \alpha_{\mathrm{CE}} \sim 0.2\text{--}0.3 \), and \( f_{\mathrm{ion}} \sim 0.2\text{--}0.4 \).

Physically, we expect a fractional recombination contribution to the final separation, as the energy must be thermalized into work to assist in ejecting the envelope \citep[e.g.,][]{demarcoFormalismCommonEnvelope2011}.
In hydrodynamical simulations, recombination is necessary for the complete ejection of the envelope after the dynamical inspiral \citep[e.g.,][]{ropkeSimulationsCommonenvelopeEvolution2023}, but the effect on the final separation is not well-understood. For some simulations, the inclusion of recombination energy can increase the final separation by 20--50\% \citep{sandCommonenvelopeEvolutionAsymptotic2020,chamandyCommonEnvelopeEvolution2020,gonzalez-bolivarCommonEnvelopeBinary2022,lauCommonEnvelopesMassive2022a}, but for others it is minimal \citep[e.g.,][]{nandezRecombinationEnergyDouble2015,reichardtImpactRecombinationEnergy2020}.
Recombination occurs in different regions of the envelope for each ion species and can thus have different thermalization efficiencies \citep{lauCommonEnvelopesMassive2022a}.
A fractional contribution is thus expected if only part of the potential (e.g., the helium) is able to perform work on the gas.
In particular, hydrogen recombination reduces the local opacity, making its thermalization less efficient \citep[e.g.,][]{sokerRadiatingHydrogenRecombination2018}. This is consistent with the idea that hydrogen recombination energy sources are mostly used to power the light curves of luminous red novae instead of exerting work on the envelope \citep[e.g.,][]{matsumotoLightcurveModelLuminous2022,chenBridgingGapLuminous2024,hatfullSimulatingStellarBinary2025,muDustFormationCommon2026}.

Hydrodynamical studies are still inconclusive about how the efficiency of the thermalization varies across donor type and companion mass, or with additional input physics such as convection and radiation transport \citep[e.g.,][]{ivanovaUseHydrogenRecombination2018,wilsonRoleConvectionDetermining2019,lauCommonEnvelopesMassive2025}.
\added{Other effects may also drastically alter common envelope outcomes, but are yet to be understood to a stage where they can be parametrised for population studies. These include accretion feedback from the main-sequence companion, \citep[e.g.,][]{chamandyAccretionCommonEnvelope2018,shiberCompanionlaunchedJetsTheir2019,lopez-camaraJetsCommonEnvelopes2022,zouJetsMainSequence2022}, and the orbital evolution from fallback of bound material after the dynamical inspiral \citep[e.g.,][]{kuruwitaConsiderationsRoleFallback2016,gagnierPostdynamicalInspiralPhase2023,nibbsWeakenedInspiralsHigh2026}}

We must emphasize that the calibration we have performed here (and all others) are dependent on the assumptions used in defining \( \lambda  \) such as the core-envelope boundary \citep{ivanovaCommonEnvelopeEvolution2013} and the underlying stellar evolution tracks used to derive them \citep[e.g.,][]{sgallettaImpactEnvelopeBinding2026}.
A stronger statistical inference on these parameters is also possible, but requires a model of the selection effects from distinguishability and survey cadence. In this study, we have chosen to not model either due to their complexity.

\subsection{Distinguishing between the \( \alpha  \)-formalism and Two-stage formalism}
\label{sec:twostage}
The Two-stage formalism only results in changes for intermediate-mass donors (Figs. \ref{fig:survival} and \ref{fig:pathways}) compared to the \( \alpha  \)-formalism, which increases survival of close and wide systems with massive companions (Fig. \ref{fig:presentday_mass}). This is not an unexpected outcome -- if the formalism only applies to HG and non-degenerate core RGB donors, then only survival in those systems should be affected.
Determining whether the Two-stage formalism is a more complete description for common envelope over the \( \alpha  \)-formalism via WD-MS PCEBs thus requires observations of G-type or earlier stars (\( M > 1\, M_{\odot} \)) at close orbital periods (1--10 days), which are challenging to distinguish observationally.

Alternatively, one can instead forward-model the observed space of naked helium stars from the surviving HG/RGB pathways shown in Figs. \ref{fig:survival} and \ref{fig:pathways}. These are observationally referred as hot subdwarf (sdOBA-type) stars \citep{heberHotSubdwarfStars2009} or stripped helium stars \citep{gotbergSpectralModelsBinary2018}.

Spectroscopy of sdOBA-type stars is typically performed assuming a canonical mass of \( 0.45 M_{\odot} \) \citep{heberHotSubdwarfStars2009}, though this may be an inaccurate assumption \citep{gotbergSpectralModelsBinary2018}. sdOBA-type stars in close binaries with masses larger than \( \sim\!0.6 M_{\odot} \) are modelled differently with Two-stage formalism as they can arise from RGB donors more massive than \( 2.25 M_{\odot} \). Properties for hot subdwarf binaries have been derived assuming a non-canonical mass by some authors \citep{schaffenrothHotSubdwarfsClose2022, dawson500PcVolumelimited2026}, but they remain to be tested against population synthesis models.

Large samples of intermediate-mass naked helium-burning stars from massive RGB donors (\( M > 8 M_{\odot} \)) are yet to be conclusively observed with measurements of both component masses and orbital periods.
While two detached systems from stable mass transfer are known: HD 966670 \citep[][]{nazeAnotherOneBH+OB2025} and HD 45166 \citep{deshmukhHighlyMagneticWolfRayet2025}, a small population (\( \sim\! 10\text{--}100 \)) of intermediate-mass naked helium stars has only recently been discovered with ultraviolet (UV) photometry in the Magellanic Clouds \citep{droutObservedPopulationIntermediatemass2023,ludwigStrippedStarUltravioletMagellanic2025}.
This population is believed to arise from envelope stripping in stable mass transfer, but no period measurements for these stars are published in the literature to confirm this.
These naked helium stars are known to be the progenitor systems of H-absent stripped-envelope supernovae (SNe Ib) \citep[e.g.,][]{podsiadlowskiPresupernovaEvolutionMassive1992,clocchiattiLightCurvesStrippedEnvelope1997}. Theoretical works have largely neglected the common envelope channel \citep{yungelsonElusiveHotStripped2024,blombergIntermediateMassStrippedStars2026} or find that it is less prevalent \citep{zapartasPredictingPresenceCompanions2017,souropanisPowerBinariesStrippedenvelope2026,ercolinoMasstransferringBinaryStars2025}. \citet{souropanisPowerBinariesStrippedenvelope2026} find that common envelope channel only comprises \( \sim \! 6\% \) of stripped-envelope SNe progenitors, but this may be higher with an increase in surviving HG and red supergiant common envelope events with the Two-stage formalism (Fig. \ref{fig:survival}).

Testing the Two-stage formalism against observations of PCEBs with naked helium stars can provide strong constraints on whether the Two-stage formalism is a more accurate description for common envelope in these mass regimes, and thus how termination of the common envelope inspiral proceeds when a star develops a significant (i.e., \( M_{\mathrm{rad}} \gtrsim 0.01\,M_{\mathrm{env}} \)) radiative region above the burning shell \citep{ivanovaCommonEnvelopeEvolution2013}.

\subsection{Adjustments to critical mass ratios and magnetic braking}
\label{sec:vary_physics}
The input physics used in our standard model described in Sec. \ref{sec:popsynth} and Table \ref{tab:params} are presented as our best-fitting models for the input physics of magnetic braking and critical mass ratios. We explore adjustments to these input physics by changing magnetic braking to the scheme used in \texttt{BSE} \citep{rappaportNewTechniqueCalculations1983,hurleyEvolutionBinaryStars2002}, critical mass ratio prescriptions to those implemented by \citet{claeysTheoreticalUncertaintiesType2014}, and the assumption of the \( L_{2} \) overflow instability in Sec. \ref{sec:qcrit}.
We show variations to Fig. \ref{fig:presentday_mass} with these adjustments in Figs. \ref{fig:yunnan_hurley_L2}, \ref{fig:tpagb_mloss}, \ref{fig:yunnan_belloni_noL2}, and \ref{fig:bse_belloni} of Appendix \ref{apd:inputphys}. We find the following:
\begin{enumerate}
	\item Altering the magnetic braking prescription only results in strong changes at CV-like periods. The saturated-disrupted prescription from \citet{belloniEvidenceSaturatedDisrupted2024} provides a slightly better match to observed CV-like PCEBs in the \( \alpha  \)-formalism than the \citet{rappaportNewTechniqueCalculations1983} braking by reducing the number of PCEBs with CV-like periods and companion masses above \( \sim \!1\,M_{\odot} \) (Fig. \ref{fig:presentday_mass} vs. Fig. \ref{fig:yunnan_hurley_L2}).
	\item Using a \( q_{\mathrm{crit}} \) that does not include the outer Lagrange point overflow criteria in giants (i.e., only the adiabatic instability \( q_{\mathrm{ad}} \)) decreases the match with observations at close, wide, and super-wide orbits (Fig. \ref{fig:yunnan_belloni_noL2}). The strong overdensity of wide and super-wide systems comprises the majority of TPAGB donors, which undergo an inspiral-driven common envelope after a long episode of non-conservative stable mass transfer that removes upwards of 90\% of the envelope (Fig. \ref{fig:tpagb_mloss}). We do not favour this model as it is both observationally inconsistent \citep[see also][for stable mass transfer pathways]{yamaguchiPopulationDemographicsWhite2025} and physically inconsistent with how we expect common envelope evolution to proceed \citep[e.g.,][]{hjellmingRapidMassTransfer1989,ivanovaCommonEnvelopeEvolution2013}.
	\item A classical \( q_{\mathrm{crit}} \) prescription \citep[e.g.,][]{hurleyEvolutionBinaryStars2002,claeysTheoreticalUncertaintiesType2014} provides similar predictions to our standard models (Fig. \ref{fig:bse_belloni}). However, the overdensity of CV-like and close PCEBs is less concentrated where observations are present, so we favour the model in Fig. \ref{fig:presentday_mass}.
\end{enumerate}
In summary, we find that magnetic braking prescriptions do not strongly affect our results, but the updated stability bounds for mass transfer lead to physically and observationally inconsistent populations. Our investigation of the mismatch found when using solely a dynamical instability criterion based on \( q_{\mathrm{ad}} \) led to the implementation described in Sec. \ref{sec:qcrit}. We would expect stronger changes from magnetic braking when modelling populations of semi-detached binaries, such as AM CVn and classical novae \citep[e.g.,][]{nelemansShortperiodAMCVn2004,kempPopulationSynthesisAccreting2021}.

\subsection{The impact of metallicity and eccentricity}
\label{sec:mh_ecc}
We only consider the Solar metallicity population as it well represents a large portion of the observed PCEB population.
Stars of lower metallicity are more compact due to a lower opacity, which should adjust when and how systems initiate mass transfer compared with a Solar metallicity population.
This compactness also increases the binding energy of the envelope and decreases \( \lambda  \) values \citep{klenckiItHasBe2021,sgallettaImpactEnvelopeBinding2026}, but this has not been explored in large detail for low and intermediate-mass stars \citep[see][]{xuBindingEnergyParameter2010,sgallettaImpactEnvelopeBinding2026}.
Mass transfer stability also shows a dependence on metallicity and this is yet to be fully explored \citep{geAdiabaticMassLoss2024}.
For these reasons, we currently do not consider metallicity variations.

We only consider circular orbits, as we expect that isolated eccentric binaries should circularize well-before mass transfer begins \citep{hurleyEvolutionBinaryStars2002}, \added{but some observations find that non-zero eccentricities are maintained in wider WD-MS systems \citep[e.g.,][]{chawlaGaiasPromiseDetect2026,yamaguchiPopulationDemographicsPostinteraction2026}. Which of these systems are products of common envelope or stable mass transfer pathways remains unknown,} but models of eccentric mass transfer remain to be tested in population studies, especially when considering higher order multiples \citep{toonenEvolutionStellarTriples2020,parkosidisRethinkingMassTransfer2026,parkosidisRethinkingMassTransfer2026a}. Variations in eccentricity for isolated binaries should also change how early systems interact by decreasing the pericenter distance.

We expect eccentricity and metallicity to be more relevant to the analysis of stable mass transfer systems, such as extrinsically enriched carbon-enhanced metal-poor (CEMP-s) stars \citep[e.g.,][]{lucatelloBinaryFrequencyCarbonenhanced2005,lugaroSprocessAsymptoticGiant2012,hansenRoleBinariesEnrichment2016} and barium stars \citep[e.g.,][]{mcclureBinaryNatureBarium1980,stancliffeFormationBariumGiants2021,levineBariumStarsMilky2026,rekhiGaiaBariumDwarfs2026}.

\section{Conclusion}
\label{sec:conclusion}
In this work, we provide the first test of two new, alternative models for common envelope against observed post-common envelope binaries (PCEBs) from the literature.
We apply the Two-stage formalism \citep{hiraiTwostageFormalismCommonenvelope2022} and the SCATTER formalism \citep{distefanoSCATTERNewCommon2023} to the low and intermediate-mass binary systems which produce white-dwarf main-sequence (WD-MS) post-common envelope binaries (PCEBs) using a binary population synthesis model.
Our main results are summarized as follows:
\begin{enumerate}
	\item \emph{The SCATTER formalism does not better predict common envelope outcomes.} The SCATTER formalism cannot predict wider PCEBs (Figs. \ref{fig:pathways} \& \ref{fig:presentday_mass}) and does not well match the observed PCEBs (Fig. \ref{fig:presentday_mass}), with parameter adjustments only quantitatively matching parts of the observed PCEB population (Sec. \ref{sec:angmom}, Fig. \ref{fig:eta_sweep}). We take this and the issues with the \( \gamma  \)-formalism as indicative of fundamental challenges with using the orbital angular momentum balance to predict common envelope outcomes \citep[e.g.,][]{webbinkCommonEnvelopeEvolution2008,ivanovaCommonEnvelopeEvolution2013}.
	\item \emph{Recombination energy is necessary to explain the observed population, but not all of it can contribute}. With a full contribution, we cannot correctly predict the orbital periods of IK Peg-like systems, instead overpredicting their presence at super-wide orbital periods in the present-day population (Fig. \ref{fig:presentday_mass}). Without it, we cannot explain the super-wide PCEBs \citep{yamaguchiWidePostcommonEnvelope2024} with any value for \( \alpha_{\mathrm{CE}} \) (Fig. \ref{fig:claeys_frec_0}). We loosely constrain pairs of values for the efficiency and the recombination contribution which can describe the population as observed (i.e., if \( \alpha_{\mathrm{CE}}\sim 0.2\text{--}0.3 \), \( f_{\mathrm{ion}} \sim 0.3\text{--}0.4 \), Fig. \ref{fig:claeys}), but stronger constraints require more simulations and observations to understand how the thermalization efficiency varies across different donor types and mass ratios.
	\item \emph{Observations and comparison to PCEBs with naked helium stars is necessary to determine if the Two-stage formalism is more accurate for describing common envelope.} Both the Two-stage and \( \alpha  \)-formalism perform equally well at predicting common envelope outcomes for observed WD-MS PCEBs, and differences are only present where observations are not (Sec. \ref{sec:obs}, Sec. \ref{sec:paramdist}, Fig. \ref{fig:presentday_mass}). Detailed observations of helium-burning PCEBs (i.e., sdOBA-type primaries) from donors more massive than \( 2\,M_{\odot} \) will provide stronger constraints on which formalism is a more accurate description for common envelope.
	\item \emph{More analysis on mass transfer stability is necessary.} We find that updated stability bounds for mass transfer \citep{geAdiabaticMassLoss2015,geAdiabaticMassLoss2020,zhangAdiabaticMassLoss2024} only result in a population which match observations when a combination of criteria based on \( L_{2} \) overflow and dynamical instability are used (Sec. \ref{sec:vary_physics} and Appendix \ref{apd:inputphys}). Using solely a dynamical instability criterion led to a nonphysical population, where the majority of wide and super-wide (\( P \sim 10\text{--}1000 \) days) PCEBs formed through an inspiral-driven common envelope after removing upwards of 90\% of the envelope mass. Similar results have been found for the stable mass transfer pathways \citep{yamaguchiPopulationDemographicsWhite2025}.
\end{enumerate}

Our results nonetheless show that the \( \alpha  \)-formalism (and more broadly, energy-based formalisms) still remain the most descriptive for all common envelope outcomes, in line with previous works \citep[e.g.,][]{davisComprehensivePopulationSynthesis2010, zorotovicPostcommonenvelopeBinariesSDSS2010, toonenEffectCommonenvelopeEvolution2013}.
Common envelope outcomes also sensitively depend on other physics than just the formalism and its parameters, including how the binding energy is parametrized and the boundary of stable mass transfer (parameterised as \( q_{\mathrm{crit}} \)). We recommend that future works holistically investigate these together when investigating common envelope uncertainties in their modelling.

Our binary population synthesis models also make several predictions that should be tested observationally. We predict abundant PCEBs with orbital periods between 1 and 100 days, containing COWDs and companions of all masses across the FGKM spectral types (\( M_{\mathrm{MS}} \sim 0.5\text{--}2\,M_{\odot} \); Fig. \ref{fig:presentday_mass}). We also expect super-wide PCEBs with G-type and earlier (\( M_{\mathrm{MS}} > 1.2\, M_{\odot} \)) and M-dwarf companions (\( M_{\mathrm{MS}} < 0.5\, M_{\odot} \)) (Fig. \ref{fig:presentday_mass}). These are both yet to be observed in populations strongly, but Gaia Data Release 4 should prove extremely helpful in discovering such systems. Data from the Rubin Observatory can also provide strong physical constraints if the recombination contribution can be related to the light curves of luminous red novae \citep{howittLuminousRedNovae2020}.

Simulations should also aim to investigate how the thermalization efficiency of recombination energy release varies with companion mass, input physics, preceding phases of mass transfer, and different evolutionary phases of the donor.
Future work can provide a more statistical constraint on the efficiency of the ejection process and the recombination contribution how these vary different donor types and mass ratios.

All of these works can provide new insights to constrain our \emph{``best guess"} for how common envelope evolution proceeds.

\section*{Acknowledgements}
The majority of this research was conducted for the completion of a Bachelor of Science (Honours) thesis at Monash University.

\added{We thank the referee for their comments which improved the quality of the manuscript.}

R.T. thanks Chiaki Kobayashi for providing the Solar neighbourhood star formation rate function, and thanks Yoshiya Mori, Thavisha Dharmawardena, and Elizabeth Iles for feedback on Figures. R.T. also thanks Alexey Bobrick and Ilya Mandel for helpful discussions.

\added{We thank Chirag Chawla and Sahar Shahaf for comments on the preprint, and Hongwei Ge for comments and providing updated values for the \( q_{\mathrm{crit}} \) tables.}

R.T. was supported by the Commonwealth through an Australian Government Research Training Program Scholarship \href{https://doi.org/10.82133/C42F-K220}{doi:10.82133/C42F-K220}.

A.J.K. wishes to acknowledge Methusalem funding from the Flemish Government through project SOUL: Stellar evolution in full glory, grant METH/24/012 at KU Leuven.

R.H. acknowledges support from the RIKEN Special Postdoctoral Research Program for junior scientists.

\added{This work was performed on the OzSTAR national facility at Swinburne University of Technology. The OzSTAR program receives funding in part from the Astronomy National Collaborative Research Infrastructure Strategy (NCRIS) allocation provided by the Australian Government, and from the Victorian Higher Education State Investment Fund (VHESIF) provided by the Victorian Government.}

\emph{Software}:
\texttt{binary\_c} \citep{izzardBinary_cStellarPopulation2023},
\texttt{SIMBAD} \citep{wengerSIMBADAstronomicalDatabase2000},
\texttt{VizieR} \citep{ochsenbeinVizieRDatabaseAstronomical2000},
\texttt{numpy} \citep{harrisArrayProgrammingNumPy2020},
\texttt{scipy} \citep{virtanenSciPy10Fundamental2020},
\texttt{astropy} \citep{astropycollaborationAstropyProjectSustaining2022},
\texttt{pandas} \citep{mckinneyDataStructuresStatistical2010},
\texttt{matplotlib} \citep{hunterMatplotlib2DGraphics2007},
\texttt{vaex} \citep{breddelsVaexBigData2018}.

\section*{Data availability statement}

\added{Parameterizations for the Two-stage formalism, \mesa{} inlists, \binc{} Python scripts, and a selection of the data used in this study is available at \href{https://doi.org/10.5281/zenodo.21636136}{doi:10.5281/zenodo.21636136}. \binc{} is a publicly available code licensed under the GNU General Public License v3.0.} More data from this paper can be made available upon reasonable request to the corresponding author \added{or by running the script in the repository}.

\begin{deluxetable*}{|l|l|}[t]
	\tabletypesize{\footnotesize}
	\tablecolumns{2}
	\tablecaption{\label{tab:params}Chosen parameters for our standard \binc{} population models.}
	\tablehead{\colhead{Parameter} & \colhead{Setting}}
	\startdata
	Simulation time                                          & 15 Gyr                                                                                                                                                                                 \\
	Metallicity                                              & \( Z_{\odot} = 0.014 \) \citep{asplundChemicalCompositionSun2009}                                                   \\
	Maximum timestep                                         & 100 Myr                                                                                                                                                                                \\
  Maximum timestep change factor & 1.2                                                                                                                                                                                    \\
	Parameter resolution                                     & \( 80\times 80\times 80 \) (512,000)                                                                                                                                                   \\
	\hline
	Primary mass \( M_{1,0} \) range                         & \( 0.8\text{--}10\,M_{\odot} \)                                                                                                                                                        \\
	Secondary mass \( M_{2,0} \) range                       & \( 0.10\text{--}10\,M_{\odot} \)                                                                                                                                                       \\
	Period \(P_{0}\) range                              & \( 1\text{--}10^{6} \,\mathrm{days} \)                                                                                                                                                 \\
	\hline
	\( M_{1,0} \) sampling distribution                      & \( \log_{10} \mathcal{U}(0.8,10) \)                                                                                                                                                              \\
	\( q \) sampling distribution                            & \( \mathcal{U}(0.1/M_{1,0},1) \)                                                                                                                                             \\
	Period sampling distribution                             & \( \log_{10} \mathcal{U}(1,10^{6} ) \)                                                                                                                                                 \\
	\hline
	\( M_{1,0} \) birth distribution                         & \citet{kroupaVariationInitialMass2001}, [0.01,150] \( M_{\odot} \)                                                                                                                     \\
	\( q \) birth distribution                               & Same as sampler                                                                                                                                                      \\
	Period birth distribution                                & Same as sampler                                                                                                                                                                        \\
	\hline
	Stellar structure algorithm                              & \citet{hurleyEvolutionBinaryStars2002}                                                                                                                                                 \\
	Initial stellar rotation                                 & 0.0                                                                                                                                                                                    \\
	Initial eccentricity                                     & 0.0                                                                                                                                                                                    \\
	\hline
  Common envelope formalism                         & \( \alpha  \)-formalism, Two-stage\tablenotemark{a}, or SCATTER\tablenotemark{a}                                                                                                                                                        \\
	\( \alpha_{\mathrm{CE}} \)                               & 0.2 (standard), or multiple (Sec. \ref{sec:discussion})                                               \\
	SCATTER \( \eta  \)                                      & fitted \( f(q_{ec}) \), Eq. \ref{eq:etafit} \citet{distefanoSCATTERNewCommon2023}                                                                                                                                      \\
	\( \lambda_{\mathrm{CE},\alpha } \)                      & \citet{claeysTheoreticalUncertaintiesType2014}                                                                                                                                                 \\
	\( \lambda_{\mathrm{CE},2S} \)                           & see Appendix \ref{sec:twostage_apply}                                                                                                                                                  \\
	\( f_{\mathrm{ion}} \)                               & 1.0 (standard), or multiple (Sec. \ref{sec:discussion})                                               \\
  \hline
	Roche-lobe overflow model                                & \citet{claeysTheoreticalUncertaintiesType2014}                                                                                                                                         \\
  Giant mass transfer stability bounds (\( q_{\rm crit} \))                                      & Yunnan \citep{geAdiabaticMassLoss2015,geAdiabaticMassLoss2020,zhangAdiabaticMassLoss2024}                                                                                                                                                 \\
	Wind Roche-lobe overflow model                                & \( q \)-dependent                                                                                                                                         \\
	Wind angular momentum mass loss                                & Spherically symmetric                                                                                                                                         \\
  Accretion limit (non-WD) & \( 10\times \tau_{KH} \)\citep{hurleyEvolutionBinaryStars2002}                                                                                                                                                                                  \\
	\hline
  White dwarf cooling and radius model & \citet{althausNewEvolutionarySequences2013} (HeWD), \citet{camisassaUpdatedEvolutionarySequences2017}                                                                                                                                                                                     \\
                                       &\& \citet{camisassaWhiteDwarfStars2025} (COWD), \citet{camisassaEvolutionUltramassiveWhite2019} (ONeWD)\tablenotemark{a}\\
	Novae and WD accretion treatment & \citet{kempPopulationSynthesisAccreting2021}                                                                                                                                           \\
	Chandrasekhar mass                                       & \( 1.38\,M_{\odot} \)                                                                                                                                                                  \\
	Minimum COWD mass for DDet SNe Ia                        & \( 0.8\,M_{\odot}\) \citep{finkDoubledetonationSubChandrasekharSupernovae2010}                                                                                                         \\
	Minimum donor \( M_{\mathrm{env}} \) for common envelope & \( 0.15\,M_{\odot}\)                                                                                                                                                                   \\
  Magnetic braking & Disrupted/saturated \citep{belloniEvidenceSaturatedDisrupted2024}\tablenotemark{a}                                                                                                                                                                                    \\
  Magnetic braking scale factor \( K \) &  50                                                                                                                                                                                    \\
	\enddata
  \tablenotetext{a}{Implemented in this work.}
\end{deluxetable*}

\pagebreak

\bibliography{novae}

\appendix{}
\section{Parametrising stellar evolution models for the Two-stage formalism}
\label{sec:twostage_apply}
\renewcommand{\thefigure}{A\arabic{figure}}
\setcounter{figure}{0}
\subsection{Where the Two-stage formalism applies}
To implement the Two-stage formalism in \binc{}, we require a numerical description of stellar structure to determine the size of the radiative and convective region of the envelope and the binding energy of each region. For this, we employ the Mount Stromlo Monash Stellar Evolution (Monash) code. The Monash code is purpose-built to accurately evolve low-and-intermediate mass stars through the to the end of the TP-AGB, and remains regularly used for stellar yields from AGB stars \citep{karakasUpdatedStellarYields2010, karakasStellarYieldsMetalrich2016}, super-AGB stars \citep{dohertySuperAsymptoticGiant2010,dohertySuperAGBStarsTheir2017}, and metal-poor AGB stars \citep{campbellEvolutionNucleosynthesisExtremely2008,campbellEvolutionNucleosynthesisExtremely2010}. Further details about algorithms in the Monash code can be found in \citet{lattanzioLowMassAsymptotic1984}, \citet{lattanzioAsymptoticGiantBranch1986}, \citet{frostNumericalTreatmentDependence1996}, and particularly \citet{karakasStellarYieldsMetalrich2016} as we use those input physics except with a Reimers mass loss rate with \( \eta =0.477 \) \citep{mcdonaldMasslossRedGiant2015}.

\begin{figure*}[t]
	\centering
	\includegraphics[width=0.8\textwidth]{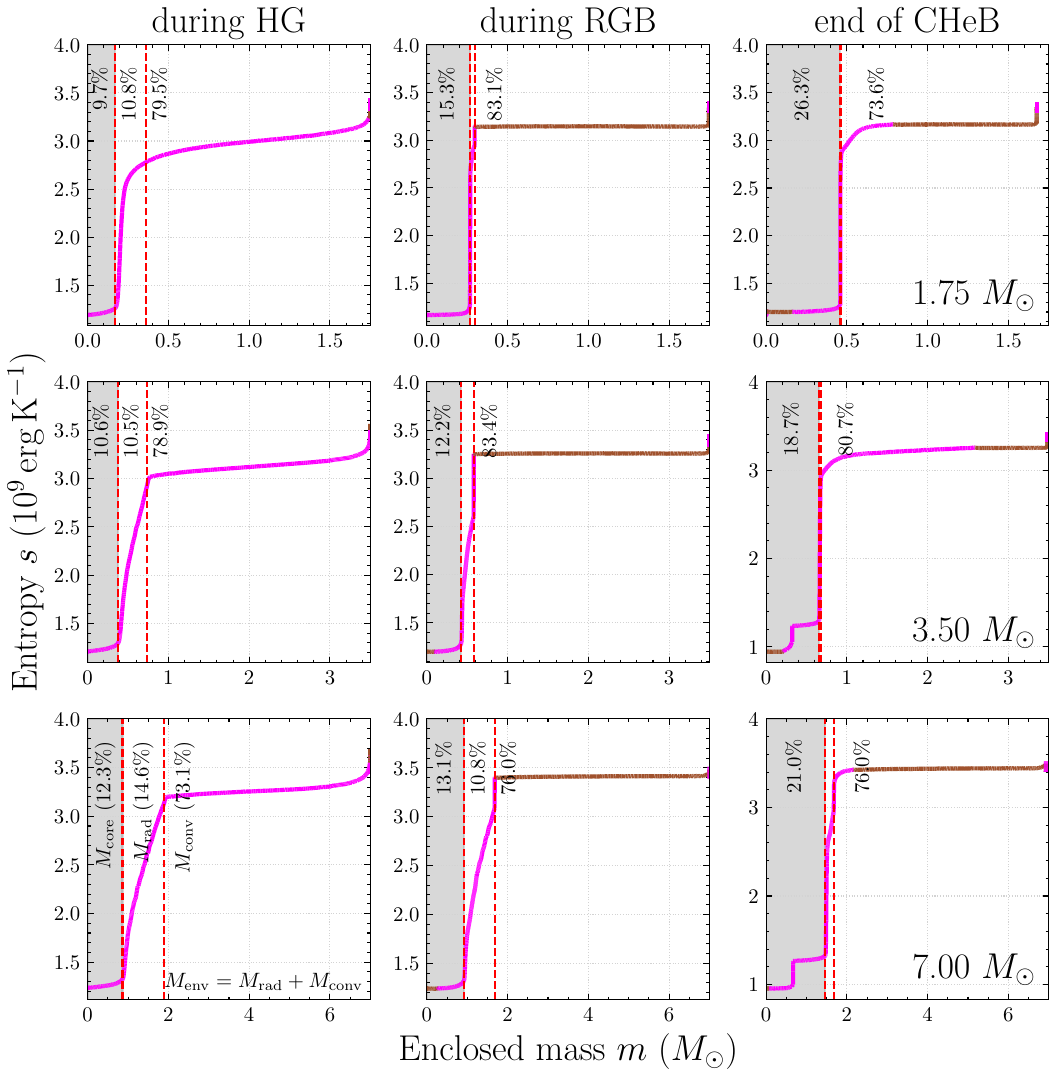}
	\caption{The ideal gas entropy profiles for three different initial masses: \( 1.75 M_{\odot} \) (top row), \( 3.5 M_{\odot} \) (middle row), and \( 7 M_{\odot} \) (bottom row). Each is shown at three different phases of their evolution (left to right): the HG, after first dredge-up on the RGB, and at the end of the CHeB phase. The fractions of the total mass the core (shaded, \added{up to hydrogen mass fraction \( X=0.01 \)}), radiative and convective/isentropic regions of the envelope are shown. We define the radiative region as the region above the core consistent of a steep increase in the entropy profile, and the convective envelope as the isentropic portion of the profile, with definitions given in the bottom left panel. }
	\label{fig:entropy_grid}
\end{figure*}

We investigated where the Two-stage formalism applies in both mass and evolutionary phase for low-and-intermediate mass stars in solar metallicity models \citep[\( Z=0.014 \),][]{asplundChemicalCompositionSun2009}.
\added{We take the separation between stages to occur in mass coordinate based on the entropy structure. We take the region where the energy budget is predictive as the (near) isentropic portion of the envelope, regardless of the local stability criteria for convection. This approach mitigates overgeneralizing the three-dimensional behaviour of common envelope evolution from the one-dimensional model \citep[see, e.g.,][]{cohenDepletionRedSupergiant2023}. We define the `core' boundary at hydrogen mass fraction \( X = 0.01 \) to allow extensions of our results upto higher masses \citep[e.g.,][]{pickerFitsConvectiveEnvelope2024}.}
Fig. \ref{fig:entropy_grid} demonstrates the evolution of this entropy profile during the evolution for a \( 1.75\,M_{\odot} \) (top row), \( 3.5\, M_{\odot} \) (middle row), and \( 7\,M_{\odot} \) star (bottom row) at three different points: the middle of the Hertzsprung gap, the end of first dredge up, and the end of core helium burning (CHeB).

For the \( 1.75\,M_{\odot} \) star, the radiative region \( M_{\mathrm{rad}} \) quickly dissipates as first dredge up penetrates deep into the stellar interior, reducing the radiative region to a very small fraction (\( <\! 3\% \)) defined by a very steep entropy gradient.
The \( 3.5\,M_{\odot} \) undergoes a similar evolution, but the shell maintains a size of \( \sim\! 4 \% \) of the total mass into the RGB. While this appears negligible, this is roughly a third the mass of the core, and thus relevant for low-mass systems.
The radiative region in the \( 7\,M_{\odot} \) starts above 14\% of the total mass and remains near 10\% throughout the evolution, unlike its lower-mass counterparts.

Across all masses, there is minimal remaining material at the end of the CHeB phase, so we conclude the Two-stage formalism does not apply along the AGB. From our models, we thus find that Two-stage common envelope evolution can only physically apply to donors \( \gtrapprox \!2.25\,M_{\odot} \), given that the radiative zone in the entropy structure is small (\( <\! 0.01~ M_{\odot} \)) below this mass. This mass is similar to the boundary where the star develops an electron-degenerate core on the RGB.

\subsection{Parametrisations of the envelope binding energy and radiative region mass}
To implement these results in \binc{}, we parametrize for both the binding energy of the convective envelope within a \( \lambda  \) prescription and the radiative region mass fraction of the envelope \( M_{\mathrm{rad}} / M_{\mathrm{env}} \) as a function of the stellar radius, fitted as
\begin{eqnarray}
	\log_{10}(\lambda)                              &=&  a_0 + a_1 x + a_2 x^2 + a_3 x^3 + a_4 x^4 + \nonumber\\
	&&a_5 x^5 + a_6 x^6, \\
	M_{\mathrm{rad}} / M_{\mathrm{env}} &=& b_0 + b_1 x + b_2 x^2 + b_3 x^3 + b_4 x^4 + \nonumber\\
	&&b_5 x^5 + b_6 x^6
	,\end{eqnarray}
where \( x = R / R_{\odot} \). We use the stellar radius as a fitting parameter as it increases in a consistent pattern across different masses during the evolution along the Hertzsprung gap and RGB, which provides leverage to describe the evolutionary point.

In our calculation of the binding energy term \( \lambda  \), we consider the thermal and radiation energy terms of the (ideal) gas, and recombination energy. We enforce all internal (and recombination) energy can be used (\( \alpha_{\mathrm{th}} = \alpha_{\mathrm{rec}} = 1 \)).

\begin{figure*}[t]
	\centering
	\includegraphics[width=0.45\textwidth]{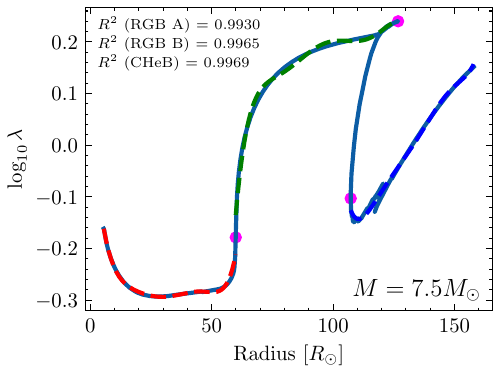}
	\includegraphics[width=0.45\textwidth]{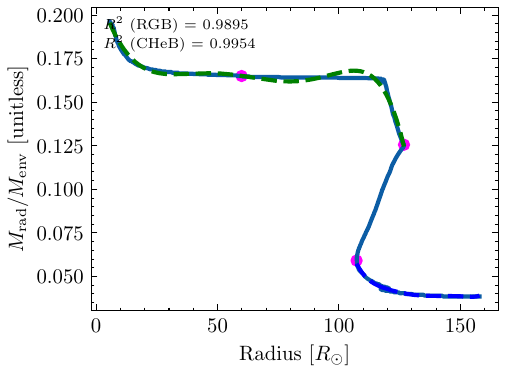}
	\caption{Examples of the polynomials fits to for the binding energy and radiative fraction of the envelope. The fits are split into the HG/RGB and the CHeB phase after re-expansion begins. \emph{Left:} evolutionary profile and fits for \( \log_{10}\lambda  \). \emph{Right:} profile and fits for \( M_{\mathrm{rad}} / M_{\mathrm{env}} \). The fit for \( \lambda  \) for the HG/RGB is split into two at the magenta point at \( \sim\!50 R_{\odot} \) to account for the steep hook shape. \( R^2 \) values for the fits are additionally denoted.}
	\label{fig:fits}
\end{figure*}

Fig. \ref{fig:fits} shows an example of the evolutionary profiles of  \( \log_{10}  \lambda  \) and \( M_{\mathrm{rad}} / M_{\mathrm{env}} \) as a function of radius, and our polynomial fits for a \( 7.5 M_{\odot} \) solar metallicity (\( Z = 0.014 \)) star. We fit separately from the start of the HG to the end of the RGB (green line) and the core Helium burning phase after re-expansion begins until helium exhaustion (blue line). We assume that any interactions in the contraction phase have been screened by the previous RGB phase, given its short lifespan. To ensure that the fit remains accurate to the evolution, we split the HG/RGB fit into two at the magenta point \( R_{\mathrm{hook}} \) for \( \log_{10} \lambda  \) (red line).

To account for stable mass transfer that results in mass ratio inversions, we complement our stellar models for the range \( 8.5\text{--}20M_{\odot} \) with models derived with \mesa{} version \texttt{r24.08.1}. We adapt the settings and control physics inlist by \citet{cinquegranaBridgingGapIntermediate2022}. This inlist is designed to produce structurally consistent evolution between both codes, and ensures the parametrizations we derive are consistent.

To calculate a \( \lambda  \) and \( M_{\mathrm{rad}} / M_{\mathrm{env}} \) for any given sampled donor mass, we interpolate the fitted values for each variable from fits to our stellar models based on mass \( M \) and the progressed radius \( R / R_{\mathrm{max}}^{\text{Monash/MESA}}\), calculated from the fits at runtime. When modelling a given binary, an \( R_{\mathrm{max}} \) for the donor is determined within \binc{} and used to determine the progressed radius the \( R / R_{\mathrm{max}} \) point for interpolation. Performing the interpolation on \( R / R_{\mathrm{max}} \) ensures the interpolated values are reflective of the current evolutionary point across both different masses and the two different stellar evolution codes used for the population synthesis models and these parametrisations \citep[e.g.,][]{sgallettaImpactEnvelopeBinding2026}.
We emphasise for other works to follow the procedure described above for using these formulae.

We provide parameters and coefficients of our polynomial fits to the binding energy \( E_{\mathrm{bind}} \) in the convective (isentropic) region of the envelope \( M_{\mathrm{conv}} \) and the radiative region mass, parametrised as a function of the envelope \( M_{\mathrm{rad}} / M_{\mathrm{env}} \) for the community.
For these fits, also attached are \( R_{\mathrm{max},\text{model}} \) values, and split values for the \( \log_{10} \lambda  \) fits at \( R_{\mathrm{hook}} \) and \mesa{} inlists for our runs.
These files are available at \href{https://doi.org/10.5281/zenodo.21636136}{doi:10.5281/zenodo.21636136}.
\section{Tests for various different input physics}
\label{apd:inputphys}
This Appendix shows Fig. \ref{fig:presentday_mass} with different adjustments to input physics for magnetic braking and critical mass ratios (Figs. \ref{fig:yunnan_hurley_L2}, \ref{fig:yunnan_belloni_noL2}, \ref{fig:bse_belloni}). Fig. \ref{fig:tpagb_mloss} shows how mass loss in TPAGB donors prior to common envelope with only a dynamical instability criterion leads to unphysical CE phases on \( <\!5\% \) of the initial envelope mass after an extended inspiral from non-conservative stable mass transfer. See Sec. \ref{sec:vary_physics} of the main text for details.
\renewcommand{\thefigure}{B\arabic{figure}}
\setcounter{figure}{0}
\begin{figure*}[t]
	\centering
	\includegraphics[width=\textwidth]{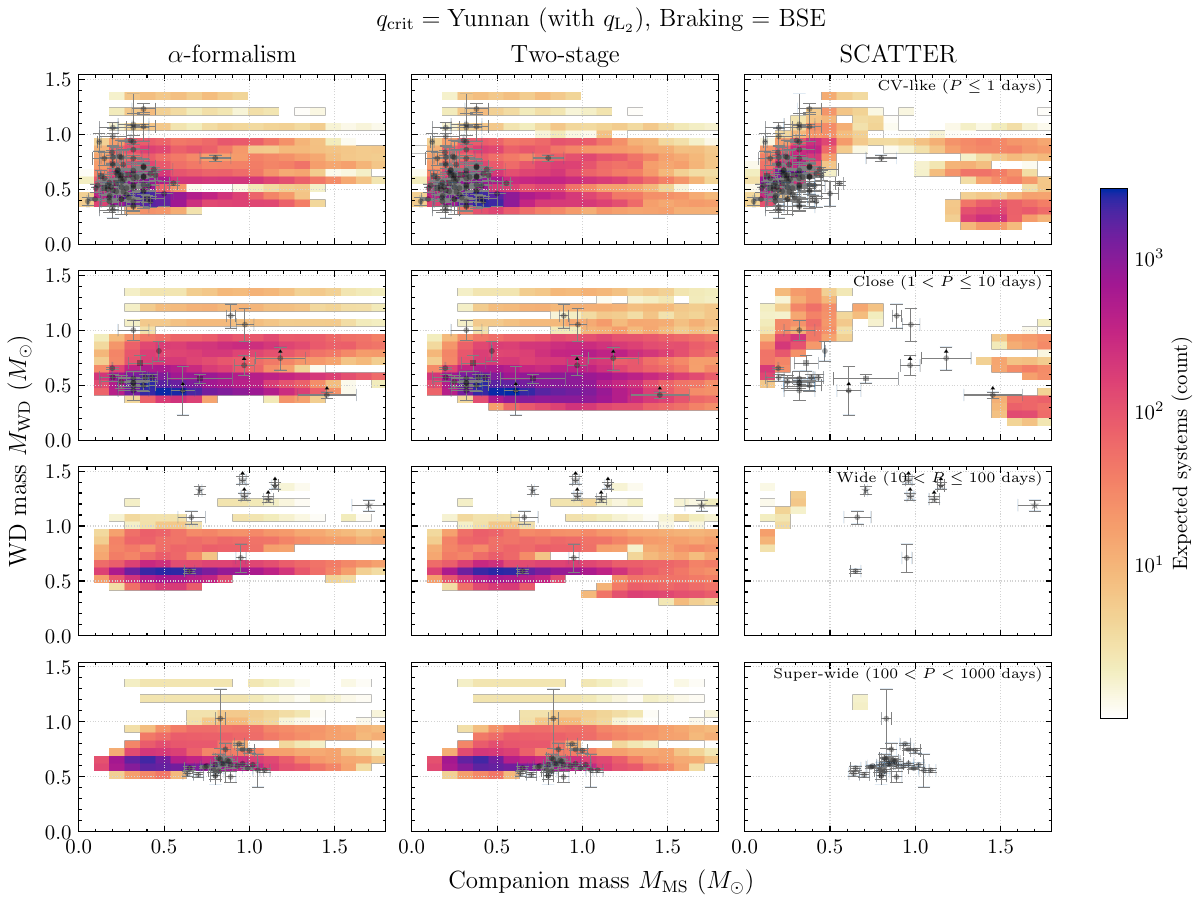}
	\caption{The mass space of present-day post common envelope binaries in our standard models, but using magnetic braking per \citet{hurleyEvolutionBinaryStars2002} (which is the prescription of \citealt{rappaportNewTechniqueCalculations1983}).}
	\label{fig:yunnan_hurley_L2}
\end{figure*}

\begin{figure*}[t]
	\centering
	\includegraphics[width=\textwidth]{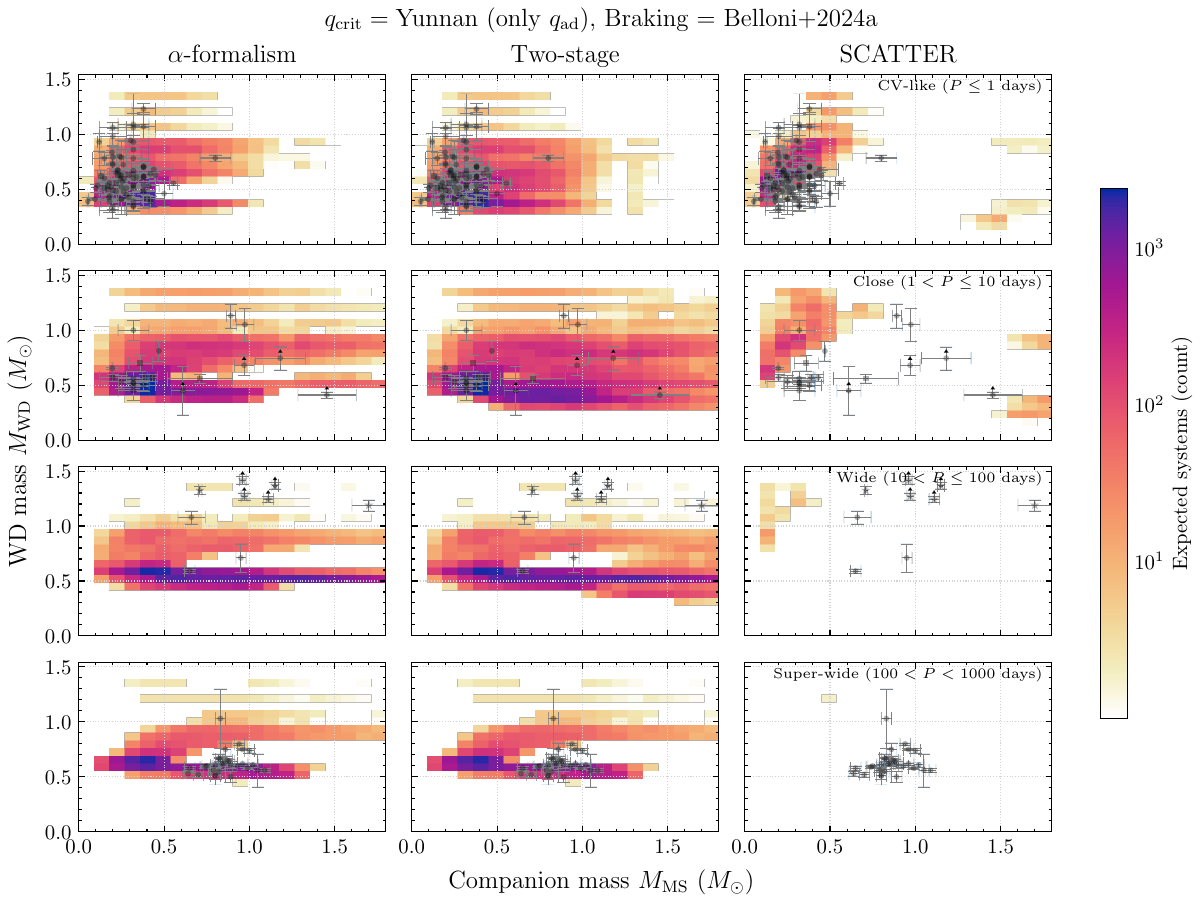}
	\caption{The mass space of present-day post common envelope binaries in our standard models, but only allowing for dynamically unstable mass transfer using \( q_{\mathrm{ad}} \) from \citet{geAdiabaticMassLoss2020}. A large number of close, wide, and super-wide systems are formed inspiral-driven common envelope with minimal envelope mass, see Fig. \ref{fig:tpagb_mloss} and \citet{yamaguchiPopulationDemographicsWhite2025}.}
	\label{fig:yunnan_belloni_noL2}
\end{figure*}

\begin{figure*}[t]
	\centering
	\includegraphics[width=\textwidth]{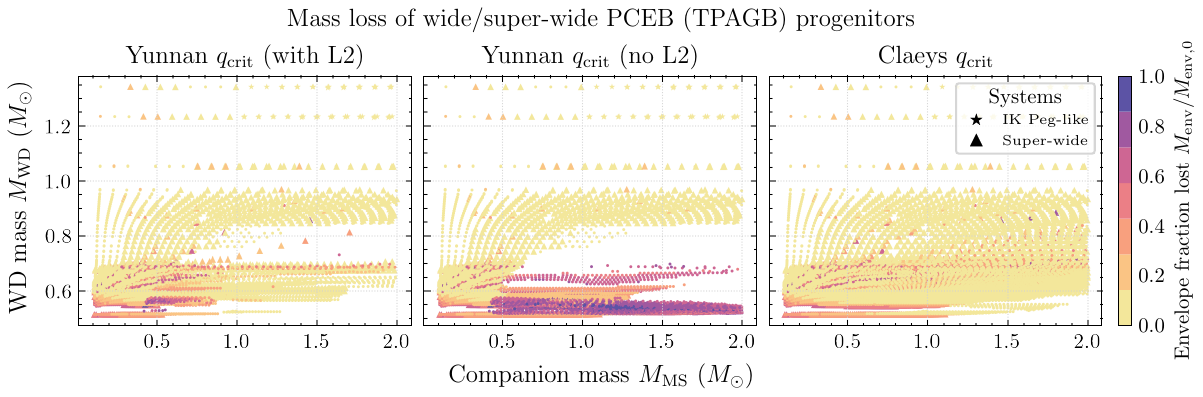}
	\caption{Mass loss in wide/super-wide TPAGB donors prior to common envelope has a dependence on \( q_{\mathrm{crit}} \). Shown are wide (\( \log P > 2 \)) systems in mass space, for each simulated point in our model. Assuming only dynamically unstable mass transfer leads to CE's on many large AGB donors which undergo a near-complete envelope stripping into a CE phase with \( \lesssim 5\% \) of the initial envelope mass, forming the majority of wide and super-wide PCEBs.}
	\label{fig:tpagb_mloss}
\end{figure*}

\begin{figure*}[t]
	\centering
	\includegraphics[width=\textwidth]{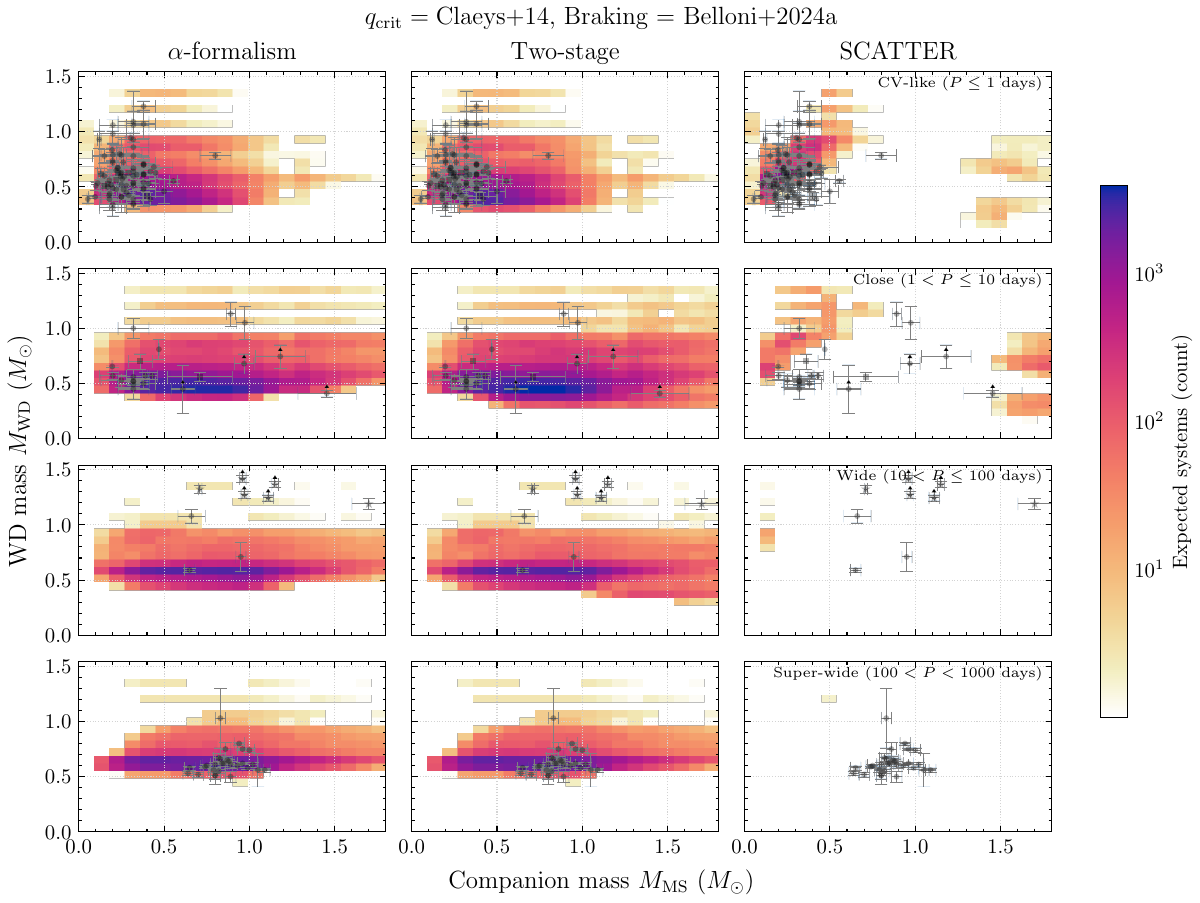}
	\caption{The mass space of present-day post common envelope binaries in our standard models, but using \( q_{\mathrm{crit}} \) values of \citet{claeysTheoreticalUncertaintiesType2014}. These values perform similarly to our standard model.}
	\label{fig:bse_belloni}
\end{figure*}

\end{document}